\documentclass[journal]{IEEEtran}
\usepackage{cite}

\ifCLASSINFOpdf
\usepackage[pdftex]{graphicx}
\graphicspath{{./Figures/}}
\else
\fi
\usepackage{color}
\newcommand{\new}[1]{#1}
\newcommand{\makefig}[3][t]{
\begin{figure}[#1]
\centering
\includegraphics[scale=1]{#2.pdf}
\caption{#3}
\label{fig:#2}
\end{figure}}

\newcommand{\makefigfull}[3][t]{
\begin{figure*}[#1]
\centering
\includegraphics[scale=1]{#2.pdf}
\caption{#3}
\label{fig:#2}
\end{figure*}}

\makeatletter
\newcommand{\Repeat}[1]{%
    \expandafter\@Repeat\expandafter{\the\numexpr #1\relax}%
}

\def\@Repeat#1{%
    \ifnum#1>0
        \expandafter\@@Repeat\expandafter{\the\numexpr #1-1\expandafter\relax\expandafter}%
    \else
        \expandafter\@gobble
    \fi
}
\def\@@Repeat#1#2{%
    \@Repeat{#1}{#2}#2%
}
\makeatother

\usepackage{lipsum}

\usepackage{upgreek}
\usepackage{xspace} 
\xspaceaddexceptions{\cite} 

\newcommand{\micron}{\ensuremath{\mathrm{\upmu m}}\xspace}
\newcommand{\nm}{\ensuremath{\mathrm{nm}}\xspace}
\newcommand{\chithree}{\ensuremath{\chi^{(3)}}\xspace}
\newcommand{\chitwo}{\ensuremath{\chi^{(2)}}\xspace}

\newcommand{\caplabel}[1]{\textbf{#1.}}

\usepackage{amsmath}
\usepackage{array}
\begin{document}
%
\def\mytitle{Silicon Quantum Photonics}

\title{\mytitle}
%
%
%

\author{Joshua W. Silverstone, ~\IEEEmembership{Member,~OSA,} Damien Bonneau, Jeremy L. O'Brien, Mark G. Thompson%
\thanks{J. W. Silverstone, D. Bonneau, J. L. O'Brien, and M. G. Thompson are with the Centre for Quantum Photonics, H. H. Wills Physics Laboratory and Department of Electrical and Electronic Engineering, University of Bristol, Bristol, BS8 1TL, UK.}
\thanks{Manuscript received February 1, 2016.}
}

%
%

\markboth{IEEE Journal of Selected Topics in Quantum Electronics, April~2016}%
{Silverstone \MakeLowercase{\textit{et al.}}: \mytitle}
%


\IEEEspecialpapernotice{(Invited Paper)}

\maketitle

\begin{abstract}
%
Integrated quantum photonic applications, providing physially guaranteed communications security, sub-shot-noise measurement, and tremendous computational power, are nearly within technological reach. Silicon as a technology platform has proven formibable in establishing the micro-electornics revoltution, and it might do so again in the quantum technology revolution. Silicon has has taken photonics by storm, with its promise of scalable manufacture, integration, and compatibility with CMOS microelectronics. These same properties, and a few others, motivate its use for large-scale quantum optics as well. In this article we provide context to the development of quantum optics in silicon. We review the development of the various components which constitute integrated quantum photonic systems, and we identify the challenges which must be faced and their potential solutions for silicon quantum photonics to make quantum technology a reality.

\end{abstract}

\begin{IEEEkeywords}
Quantum optics, silicon photonics
\end{IEEEkeywords}

%
\IEEEpeerreviewmaketitle

\section{Introduction}
\IEEEPARstart{T}{he} unique behaviour of quantum systems, exhibiting such properties as superposition and entanglement, can be harnessed to process, transmit, and encode information. Quantum information science promises to revolutionize information technology,  including the communication \cite{Bennett:1992jx, Ekert:1991kl}, processing \cite{Ladd:2010kq, AspuruGuzik:2012ho}, and collection \cite{Giovannetti:2011jka} of information. Photons---single quanta of light---and optics are at the forefront of this  quantum revolution \cite{OBrien:2009eu}.

 \makefigfull{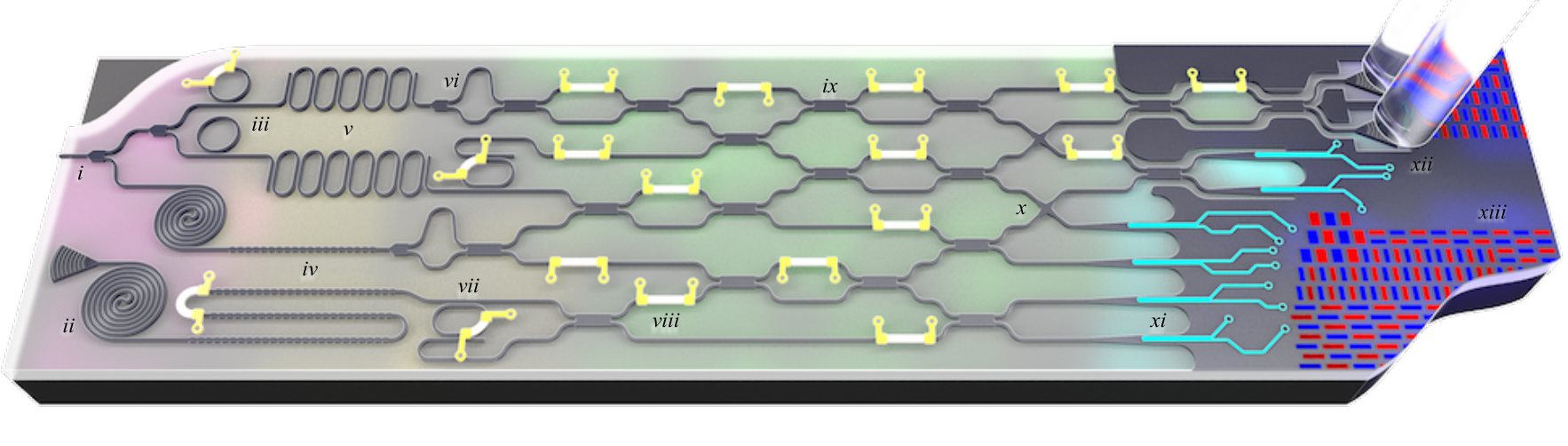}{Mock-up of a quantum photonic device, showing how various components might fit together, and what those components might look like. Regions of the chip are color coded (color online). From left to right: photon sources (magenta), pump-removal filters (yellow), passive and active optics (green), single-photon detectors (cyan), and control and feedback electronics (blue). Labels indicate: \caplabel{\emph{i}} pump input and splitter, \caplabel{\emph{ii}} spiralled waveguide photon-pair source, \caplabel{\emph{iii}} ring resonator photon-pair source, \caplabel{\emph{iv}} Bragg reflector pump removal filter, \caplabel{\emph{v}} coupled-resonator optical waveguide (CROW) pump removal filter, \caplabel{\emph{vi}} asymmetric Mach-Zehnder interferometer (MZI) wavelength-division multiplexer (WDM), \caplabel{\emph{vii}} ring resonator WDM, \caplabel{\emph{viii}} thermal phase tuner, \caplabel{\emph{ix}} multi-mode interference waveguide coupler (MMI), \caplabel{\emph{x}} waveguide crossing, \caplabel{\emph{xi}} superconducting nanowire single-photon detector (SNSPD), \caplabel{\emph{xii}} grating-based fibre-to-chip coupler, and \caplabel{\emph{xiii}} control and logic electronics.}

Integrated optics has unlocked new levels of scale and performance in quantum optics. Early demonstrations used glass-based waveguides to achieve high-fidelity on-chip  quantum interference and quantum logic operations \cite{Politi:2008jg}, multi-particle quantum walks \cite{Peruzzo:2010tq}, the reconfigurable generation and manipulation of  entanglement \cite{Shadbolt:2012bw}, and bosonic sampling \cite{Spring:2013do, Crespi:2013fu, Tillmann:2013jv}. 
However, these glass-based approaches to 
integrated quantum photonics are already reaching the limitations  of scalability and circuit complexity---with even simple circuits requiring 10-cm devices, and with little hope for scaling up in complexity and functionality. Silicon photonics promises to allow quantum optics to scale to new heights. Many of the qualities which allowed silicon to prevail as the dominant material of electronics are the same qualities which afford it a second glance optically. Silicon is abundant, has excellent thermal conductivity, is mechanically robust, can be made unimaginably pure, can be doped, and---most importantly---has an inert  oxide with which it forms high-quality interfaces. Silicon waveguides are formed by etching the device layer of a silicon-on-insulator (SOI) wafer, with confinement provided by the buried oxide (BOX) underneath and a capping oxide above.  Wafers with a 220-nm thick silicon device layer and a 2-\micron thick BOX have  become standard for photonic structures for use in the 1.55-\micron telecommunications C-band. These wafers yield tightly-confining waveguides with a  typical cross-section of $220 \times 450\ \nm^2$,  with bend radii around 5~\micron and an associated large integration density \cite{Sun:2013kl, Han:2015jh}.

The promise of an optical interconnect, whereby communications between electronic processor cores and between processor cores and memory \cite{Sun:2015gg} are handled optically---increasing range and bandwidth---has strongly motivated  the development of silicon photonics so far. For the same reasons of scale,  functionality, and manufacturability, silicon photonics may give us a crucial edge in building future photonic quantum devices.

Each quantum application---communication, sensing, computation, etc.---places its own set of requirements on the underpinning photonic technology, but these applications also have many requirements in common. Fig.~\ref{fig:overview} shows how a generic silicon quantum photonic device might fit together. All quantum applications  need \emph{a source of single photons}, with various photon statistics. All applications  then require those photons to be manipulated---either to encode information, to make measurements, or to prepare a particular quantum state---using \emph{passive optics}  to coherently mix photonic modes, and \emph{active optics} and \emph{optical delay lines} to reconfigure those modes on the fly. Once the photons have performed their useful evolution, we must always extract some classical information about their resulting state,  and we do this using \emph{detectors sensitive to single photons}.  To prevent these detectors from saturating in the presence of bright pump  fields---needed by many quantum light sources---we require very \emph{high extinction filters}, to separate single-photon signals from these bright fields. Finally, many applications would benefit from efficient  and broadband \emph{fibre-to-chip couplers}, to aid in implementing those functions more naturally accomplished off-chip, such as delay lines, and the provision of  high-power pump lasers. All these building blocks have now been demonstrated separately,  and with mixed performance, by the silicon quantum photonics community. The challenge  for future work, then, is clear: to improve the performance of individual blocks to levels sufficient for each quantum application, and to integrate these high-performance  blocks on a common silicon substrate.

Silicon is a powerful workhorse for achieving the goals of quantum photonics, but is not without its limitations. When it comes to component-level development and integration, the silicon material system is pre-eminent, with exponential improvements in microelectronics in both development and integration for the better part of the past century. Optically, silicon benefits from a strong third-order (\chithree) nonlinearity, whereby the material refractive index varies with the optical intensity.  This enables a wide range of devices, from photon sources (\ref{sec:sources}) to  photo-optic switches (\ref{sec:switches}). On the other hand, nonlinear two-photon  absorption (TPA) is the Mr. Hyde to the third-order nonlinearity's Dr. Jekyll. It presents a challenge for high-powered nonlinear optics, leading to parasitic free-carrier  and thermal effects. The small size and high confinement of silicon waveguides makes their properties vulnerable to small differences in fabrication---most loss in SOI  waveguides, for instance, arises from the roughness of their etched sidewalls, rather  than from any intrinsic absorption effect \cite{Morichetti:2010hp, Lee:2001kc}. Despite these challenges, the community has made remarkable progress. We review this progress in section~\ref{sec:review}.


\section{Quantum Photonic Devices}
\label{sec:review}

\subsection{Photon Sources}

\label{sec:sources}

Many techniques exist for the production of quantum light and single photons \cite{gr-njp-2004}; which one is most suitable depends on the application.  One single photon is well approximated by attenuated laser light, for example, where a pulse's chance to contain more than one photon is  attenuated faster than its chance to contain only one. This fact is widely used in quantum key distribution systems\cite{sc-rmp-81-1301, Sibson:2015uv} which encode information on single photons.  True single photons, on the other hand, can be obtained directly from atom-like emitters, such as trapped ions or quantum dots\cite{Ei-aip-82-071101}.  Tremendous progress has been made over the past few years in improving the specifications of true single-photon emitters.   They remain, however, hard to manufacture and do not offer much spectral flexibility. Notably, quantum dots have recently been integrated with other on-chip   quantum optics\cite{Reithmaier:2015ju}, and have obtained high levels of single-dot pulse-to-pulse indistinguishability\cite{Wei:2014dg}, bringing them nearer to system-level integration. 

Pairs of photons can be spontaneously created in silicon, owing to its nonlinear optical properties.  Two photons from a bright pump laser can be spontaneously shifted to different wavelengths via an elastic process called spontaneous four-wave mixing (SFWM).  These two photons, often referred to as signal and idler, are quantum correlated, preserving the phase of the original pump.  Due to silicon's crystallinity, noise from spontaneous Raman scattering is localized at exactly 15.6 THz from the pump frequency, making this noise easier to engineer  against than in other systems, like silica fibre and chalcogenide glasses \cite{xi-apl-98-51101}.  If generated in a monomode waveguide, the signal and idler photons will emerge in that waveguide's single transverse mode, eliminating mode matching and photon collection issues. Silicon SFWM sources also integrate naturally with passive optics, as they are formed in the same silicon patterning and etching steps.

Compared with the true single photon emitters, SFWM sources have the following drawbacks: the quantum state produced is a squeezed state which can approximate  a photon pair only if the pump laser is relatively weak. This makes the generation process probabilistic\footnote{Several statistics, from thermal to Poissonian, may arise, depending on the conditions of operation\new{\cite{wo-pra-80-053815, Tapster:1998cv}}.}: each time a pump pulse is used to seed the source, either no pair is produced, a single pair is produced, or multiple pairs are produced (a noise source in many quantum information tasks). An ideal photon-pair source is operated with thermal statistics \cite{wo-pra-80-053815} which suppress correlations between signal and idler photons. These correlations are responsible for a decreased purity when implementing heralded single-photon sources.

\new{Assuming an ideal photon-pair source, which emits into only two spectral modes, the probability to obtain $n$ pairs per pulse is given by\cite{wo-pra-80-053815, Gerry:2005hc, Helt:2012jj}
\begin{equation}
  p_n = (\text{sech}|\xi| \cosh^n{|\xi|})^2 ({\Delta \nu_c}/{\Delta\nu})^n
  \label{eq:thermal_proba}
\end{equation} 
where $\xi$ is the squeezing parameter and $|\xi|^2$ is the pair generation probability when the pump power is low, $\Delta \nu$ is the full emission bandwidth, and $\Delta \nu_c$ is the collection bandwidth ($\Delta\nu_c \leq \Delta\nu$).} The probability to emit a single pair reaches a  maximum around 25\%.

This type of source is not scalable---the probability for $N$ sources to each emit exactly one pair decreases exponentially with $N$.  However, these sources can be used to \emph{construct} near-deterministic single-photon sources, with the addition of single-photon detectors, a switching network, and suitably fast control electronics \cite{Migdall-pra-66-053805}---in a so called `source multiplexing' configuration.

\makefig{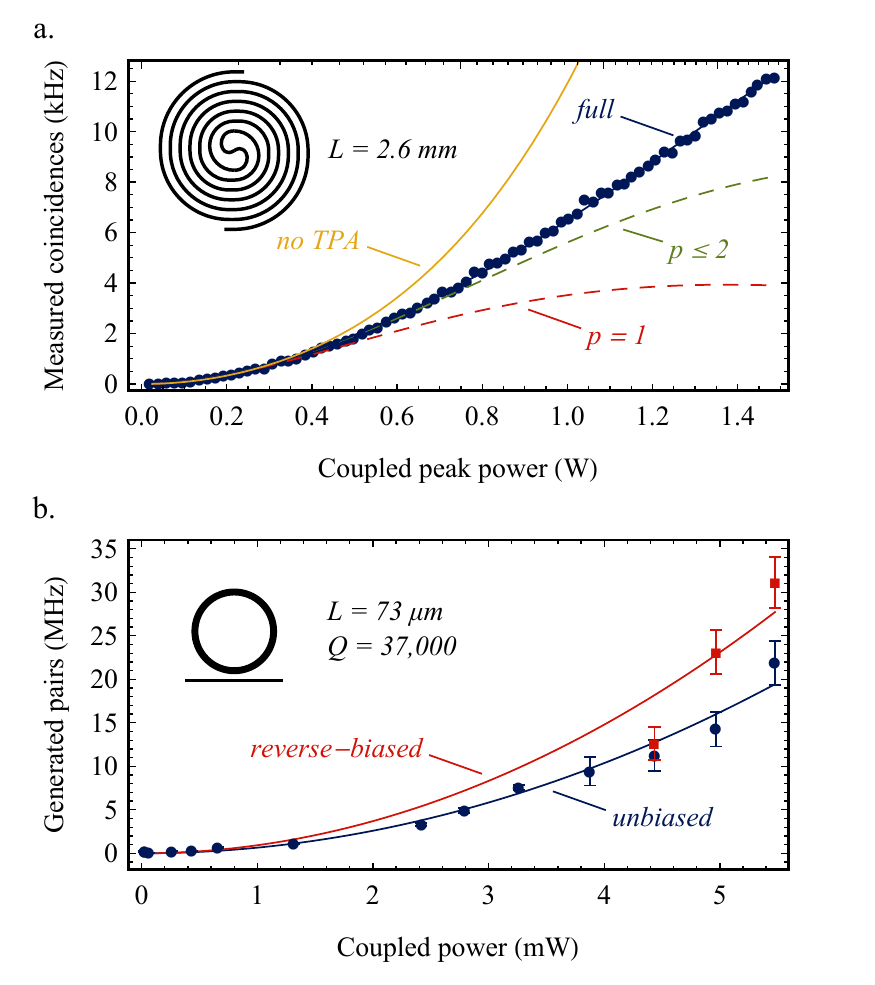}{\new{Sample measured pair generation data.} \caplabel{a} Measured pair generation rate in a 2.6mm long spiral as a function of the injected pulsed pump power (blue dots). The data is fitted with a model including multi-pair emission and non-linear absorption. The two dotted curves represent incomplete models (both fitting well at low power) which accounts respectively for single-pair emission (red) and up to two-pair emission (green). The yellow curve show the predicted pair detection rate if TPA is not included in the model. \caplabel{b} Pair generation rate in a silicon ring resonator as a function of the injected CW pump power in the absence (blue) and in the presence (red) of reverse bias to mitigate FCA \cite{Engin:2013fb}.}

SFWM depends on several parameters. The bandwidth of the generated photons is governed by energy conservation, the pump bandwidth, phase matching  (dependent on the waveguide dispersion), and the presence or absence of a cavity \cite{ag-ol-31-3140}. Since SFWM consumes two pump photons, its efficiency (when the pump is relatively weak) grows quadratically with pump power, as shown in Eq. \ref{eq:ppp}, below. Provided the power is kept low enough to neglect multi-pair emission, the pair generation rate is \cite{Husko:2013bh}
\begin{equation}
  |\xi|^2 \approx \gamma^2 L_\mathrm{eff}^2 P^2 \Theta^2.
  \label{eq:ppp}
\end{equation}
\new{$\gamma = {n_2 k_0}/{A}$ is the nonlinear parameter, with $n_2$ the nonlinear refractive index, $k_0$ the vacuum wavenumber, and $A$ the effective mode area \cite{Rukhlenko:2012hv}; $L_\mathrm{eff} = (1 - e^{-\alpha L})/\alpha$ is the effective interaction length accounting for linear loss $\alpha$; $P$ is the pump power; $\Theta = \mathrm{sinc}(\Delta k L)$ is the phase matching coefficient, with $\Delta k$ being the momentum mismatch between the four fields. }

 While SFWM is efficient and useful in silicon waveguides, other less useful nonlinear phenomena also occur. Foremost among these is two-photon absorption (TPA), which increases the single-photon propagation loss both directly, via cross two-photon absorption (XTPA) where one single photon and one pump photon are absorbed together to excite an electron, and indirectly, by being absorbed by one of the free carriers generated in this way (FCA). These effects have been modelled and carefully observed in straight waveguide pair generation experiments \cite{Husko:2013bh}. The pair generation rate is no longer quadratic with the pump power, but scales as
\begin{equation}
|\xi|^2 \overset{\text{TPA}}{\longrightarrow} \gamma^2 A^2 \alpha_2^{-2} \log(1+\alpha_2 P L_\mathrm{eff}/A) \Theta^2
\end{equation}
where $\alpha_2$ is the TPA coefficient. This leads to a saturation in the pair generation rate due to TPA. This model is used to describe measured data in Fig. \ref{fig:sfwmrates}a.
 

\makefig{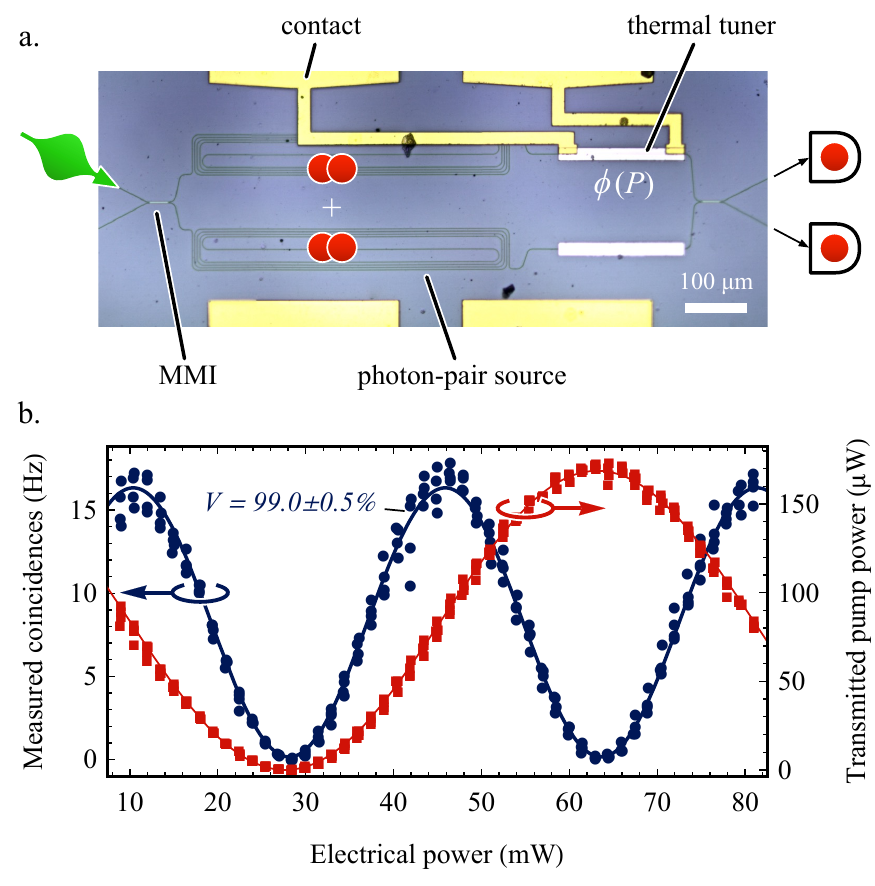}{\caplabel{a} A bright beam is injected and split on the first MMI, then pumps simultaneously the top and bottom spiral arms of the MZI source, generating a pair in superposition of being on the bottom of top arm which interferes on the last MMI. By varying the MZI phase, the two-photon output state can be tailored to bunch in one output or split between two outputs. \caplabel{b} Two-photon fringe obtained when monitoring the two-photon coincidences between two different outputs.}

Resonators improve several aspects of the photon-pair source. First, they typically have a much smaller footprint than straight- or spiralled-waveguide sources, allowing a higher integration density. Second, they enhance the brightness by increasing the circulating intensity of the resonant pump, thereby reducing the pump power required to achieve a given pair generation rate. Third, they affect the joint spectral emission of the photon pair, forcing the signal and idler to each have a spectrum centered on a resonant cavity mode. By adjusting the pump bandwidth, one can also ensure that pairs of emitted photons are spectrally separable (uncorrelated) \cite{lgh-ol-35-3006}, thus making the resonant source a good candidate for the heralding of single photons. \new{This is supported by measurements of the joint spectral intensity of silicon ring sources \cite{Silverstone:2015cl, Kumar:2014gb}, and by measurements of the correlation function of pairs generated in silicon microdiscs \cite{Lu:2016wu} and in hydex and silicon nitride rings \cite{Reimer:2015bv, Ramelow:2015uf}. }

\new{In a resonator, the phase matching condition is naturally satisfied, as cavity resonances are evenly spaced in wavenumber.} Energy conservation still applies, however, and a source cavity must be triply resonant: for the consumed pump, and for the generated signal and idler photons. This is typically the case for neighbouring resonances in silicon rings, for which the dispersion is roughly linear and so the cavity resonances are roughly evenly spaced in frequency as well as in wavenumber\footnote{As a result, energy is conserved ($2\nu_p = \nu_s + \nu_i$), in addition to momentum, which is always conserved for equally spaced resonances ($2k_p = k_s + k_i$).}. On the other hand, operating silicon resonators requires extra care: both TPA and FCA have associated phase effects which can modify the cavity resonances as a function of the intra-cavity power. The exact impact of these phenomenon on the biphoton joint spectral amplitude, and its potential impact on the purity of a heralded single photon source is a topic of current research.

Photon-pair generation in a silicon nanowire was first reported by Sharping \emph{et al.} in 2006 \cite{Sharping:2006tv}. Since then, many SFWM experiments have been reported: in strip waveguides \cite{Sharping:2006tv, Harada:2011cw, Harada:2008iy, Clemmen:2009tc}, with control over polarization \cite{Matsuda:2012dm} (Fig. \ref{fig:sources_bundle}d), in a phase stable interferometer with \cite{Silverstone:2014fu} (Fig. \ref{fig:rhom}) and without \cite{Olislager:2013gi} (Fig. \ref{fig:sources_bundle}e) tunable elements. Multiple types of resonator-based sources have also been investigated: single-ring resonators \cite{Azzini:2012io, Grassani:2015ft, Wakabayashi:2015cl, Preble:2015tb, Harris:2014kj, Engin:2013fb, Silverstone:2015cl} (Fig. \ref{fig:sources_bundle}c), including studies of the effect on reverse bias PIN junctions for mitigating FCA\cite{Engin:2013fb, sa-apl-107-131101} (Fig. \ref{fig:sfwmrates}b); ring resonators in a self-locking double-bus configuration\cite{Reimer:2014ev} (Fig. \ref{fig:sources_bundle}j), \new{in a quantum splitter configuration \cite{He:2015ej} (Fig. \ref{fig:sources_bundle}i),} and in coupled-resonator optical waveguides (CROW) \cite{Kumar:2014gb, Davanco:2012et} (Fig. \ref{fig:sources_bundle}f); high-Q microdiscs \cite{Jiang:2012wt, Lu:2016wu, ro-apl-107-041102} (Fig. \ref{fig:sources_bundle}h); photonic crystal waveguides \cite{Xiong:2011gt, Collins:2013eu} (Fig. \ref{fig:sources_bundle}a), and coupled cavities\cite{Matsuda:2013cv, Takesue:2014ic} (Fig. \ref{fig:sources_bundle}b); and in one-dimensional photonic crystal resonators\cite{Azzini:1563418} (Fig. \ref{fig:sources_bundle}g).

\makefig{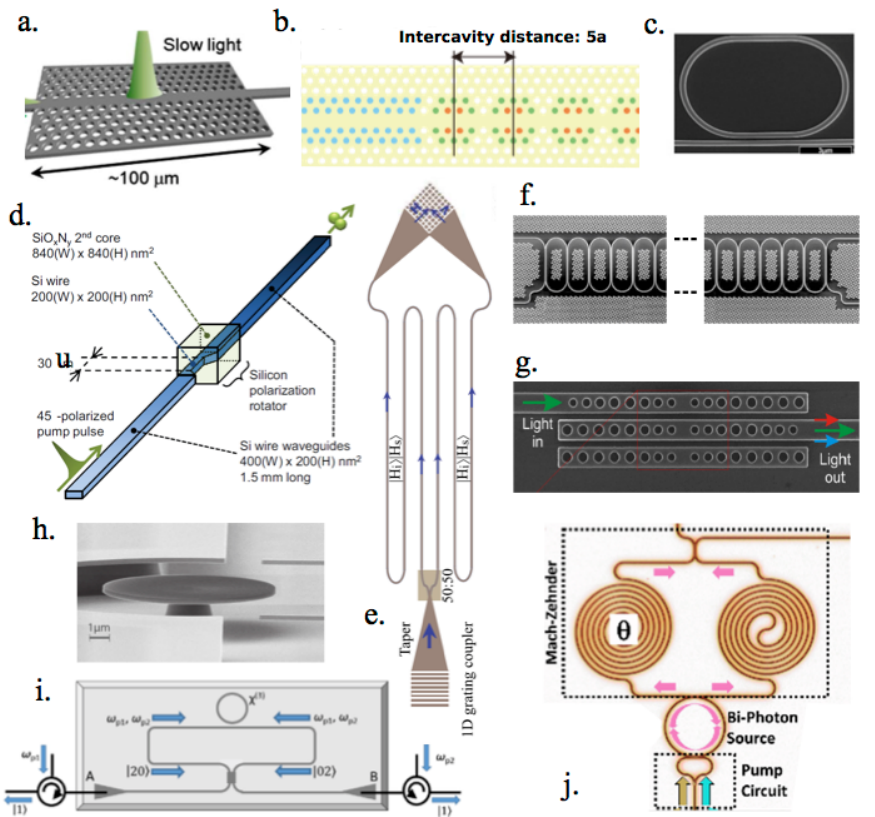}{Examples of structures used for the generation of photon pairs:
    \caplabel{a} photonic crystal waveguide,
    \caplabel{b} coupled photonic crystal cavities,
    \caplabel{c} ring resonator,
    \caplabel{d} waveguide with polarisation rotator,
    \caplabel{e} two-path source with a 2D-grating combiner,
    \caplabel{f} coupled ring resonator optical waveguide,
    \caplabel{g} coupled one-dimensional photonic crystal resonators,
    \caplabel{h} microdisc resonator,
    \caplabel{i} bidirectionally, non-degenerately pumped all-pass ring, and 
    \caplabel{j} \new{bidirectionally, non-degenerately pumped add-drop ring.}
    See main text for references.
 }

\subsection{Passive Optics}

\makefig{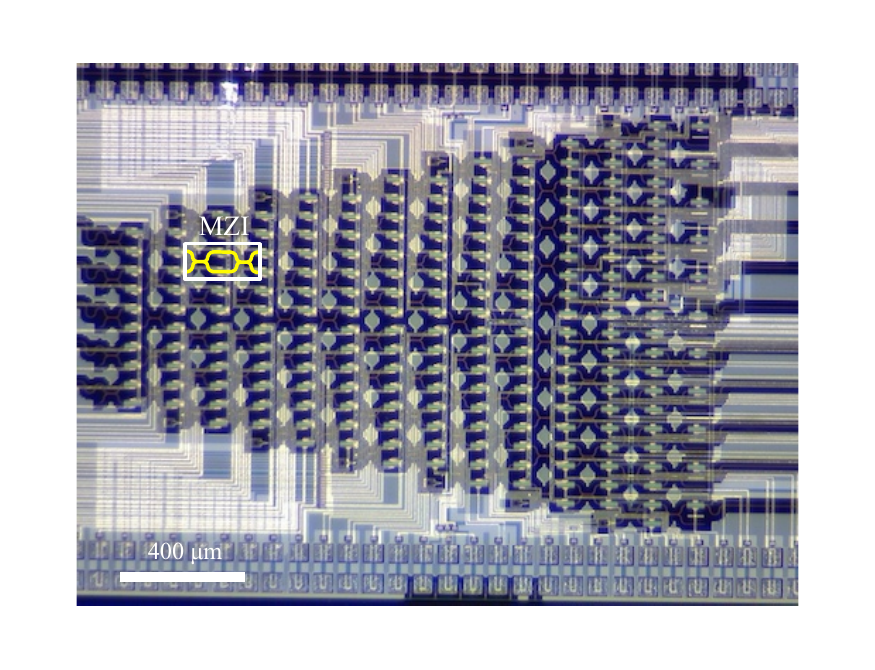}{Programmable nanophotonic processor, from reference \cite{Harris:2015ux, Steinbrecher:2015ej}. A Mach-Zehnder interferometer (MZI) is highlighted. The device contains 56 such interferometers, and 213 phase tuners to control their interferences. Image credit: N. C. Harris.}

\makefig{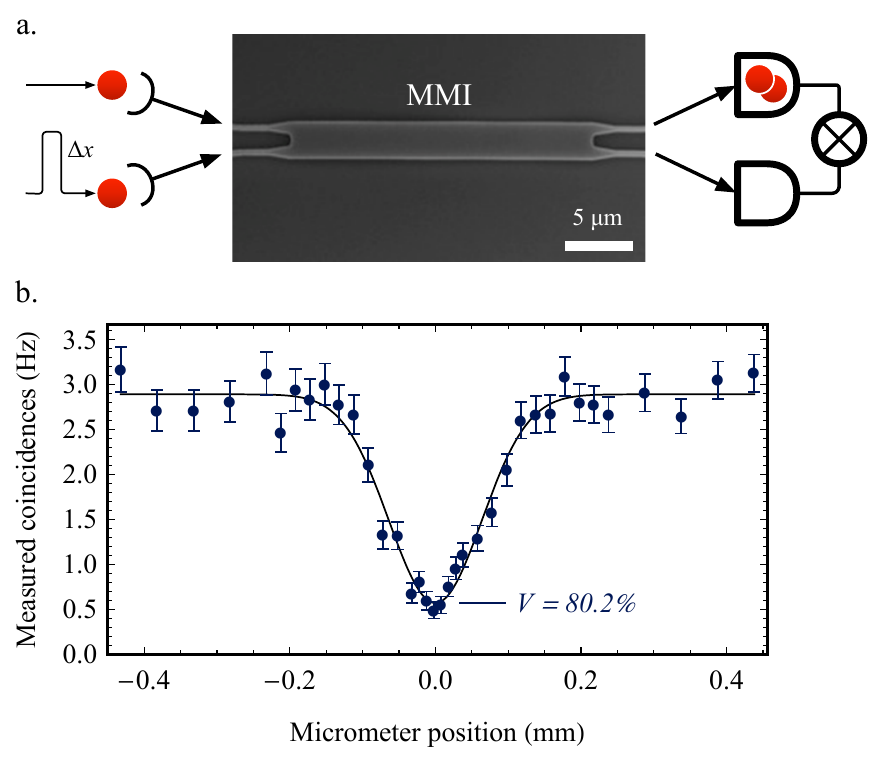}{\caplabel{a} Schematic of a two-photon quantum interference experiment. Two photons are launched in a silicon MMI. A tunable delay line at the input allows for varying the relative arrival time (delay) between the two photons. Each output port of the MMI is monitored with a single photon detector and correlated detection events (coincidences) are recorded. \caplabel{b} Number of coincidences as a function of the delay. A drop in the coincidences is observed as the wavepackets of the two photons overlap, leading to destructive interference in the state where they exit different output ports.}

Once a quantum state of light has been generated, quantum information is mapped onto and off of this state coherently, using passive, linear optics. In some applications, these passive optics are relatively simple; in others, they form intricate nested interferometers. QKD can function with as few as one Mach-Zehnder interferometer at the transmitter and one at the receiver, to share a secure quantum key. Linear optical quantum computation, on the other hand, is likely to require thousands or millions of passive optical elements \cite{Li:2015ue}.

The most complex quantum photonic device reported to date consists of 30 directional couplers enclosing 30 thermal tuners on 6 modes, etched into glass waveguides on a single $40 \times 100~\mathrm{mm}^2$ die \cite{Carolan:2015fba}. Boson sampling, whereby a large multiphoton state is measured after propagating through a large interferometer, thereby sampling the bosonic distribution, may represent the first opportunity to show that quantum devices are classically unsimulable \cite{Aaronson:2011ho}. Several groups have reported cm-scale laser-written waveguide implementations of the boson sampling experiment, with 10, 12, and 72 waveguide couplers (refs. \cite{Spring:2013do} and \cite{Crespi:2013fu}, \cite{Tillmann:2013jva}, and \cite{Bentivegna:2015ia}, respectively). These impressive results start to run up against the limits of what is possible on a single substrate. Furthermore, none of these laser-written devices incorporated electro-optic tuning elements. A higher component density is the only way to grow these devices further, and, though they excel at producing low-loss passives, these glass-based integrated optics lack facilities for doing much else.

Silicon waveguides offer solutions to these problems, with unmatched component densities, and added functionalities based on silicon’s semiconducting nature. The quantum optics community has already begun to exploit the density and reproducibility of silicon’s passive structures. A silicon photonic device with 112 multi-mode interferometers (MMIs) and 213 silicon-based tuners (Fig. \ref{fig:mitpnp}) has recently been demonstrated by Harris, Steinbrecher \emph{et al.} \cite{Steinbrecher:2015ej, Harris:2015ux}, with a scale substantially exceeding those of the aforementioned glass-based devices. Experiments on this device have so far used only bright light, though its design is for single photons.

Quantum optics, based on the passives available in silicon photonics, has advanced rapidly in recent years. After the generation of photon pairs in silicon waveguides was demonstrated
in 2006 \cite{Sharping:2006tv}, the first quantum interference in silicon photonics was demonstrated in 2012. Bonneau \emph{et al.} used $2\times 2$ MMIs to construct a classic Hong-Ou-Mandel (HOM, \cite{Hong:1987vi}) interferometer (shown in Fig. \ref{fig:hom}), and a MZI, and used these two experiments to show quantum interference visibilities exceeding 80\% \cite{Bonneau:2012gx}. Subsequently, Xu \emph{et al.} extended this work to directional couplers, showing a HOM dip with a visibility of 90.5\% \cite{Xu:2013jna}. Silverstone \emph{et al.} showed that SFWM photon-pair sources were straightforward to integrate with other passive optics in 2013 \cite{Silverstone:2014fu}, obtaining on-chip quantum interference visibilities approaching unity (see Fig. \ref{fig:rhom}). This development lead to the demonstration of entangled single-qubit logic in 2015 \cite{Silverstone:2015cl}, again with photons generated on chip, but this time powered by ring-resonator-based photon-pair sources. Also in 2015, Silverstone, Santagati \emph{et al.} demonstrated multi-qubit logic applied to on-chip-generated entangled photons \cite{Silverstone:2015tl}, in a device with 21 passive elements, 16 thermal tuners, and 4 photon-pair sources (producing one pair in superposition between them). Silicon photonics offers passive optical components with high density, high yield, high performance, and which can be combined with extra on-chip functionalities. These robust components will form the backbone of any future silicon quantum photonic computer, communications system, or metrology device.

\makefig{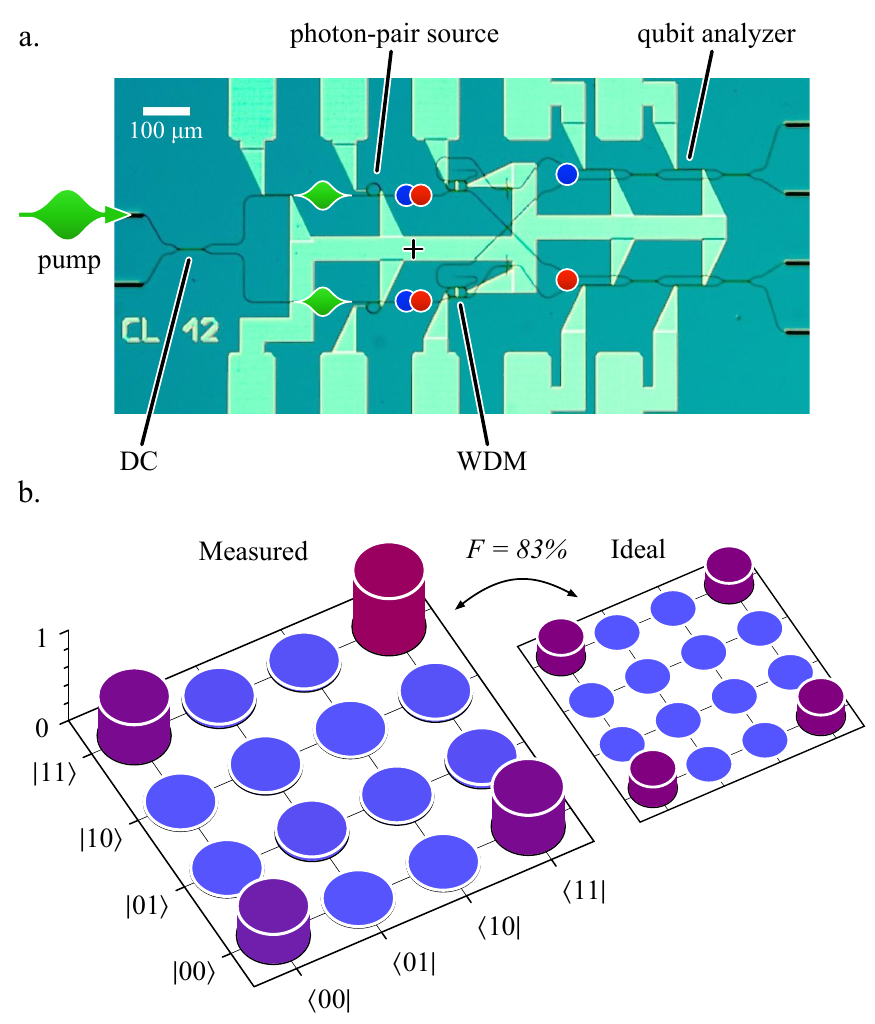}{On-chip Bell state generation and analysis, from reference \cite{Silverstone:2015cl}. \caplabel{a} Optical micrograph of the device, with components annotated; directional coupler (DC), wavelength division multiplexer (WDM). A bright pump generates non-degenerate photon pairs in superposition between two resonant SFWM sources. These pairs are reconfigured and analyzed on-chip in the path qubit basis. \caplabel{b} An on-chip-reconstructed two-qubit entangled state. Image credit: M. J. Strain.}

\subsection{Single-Photon Detectors}

\makefig{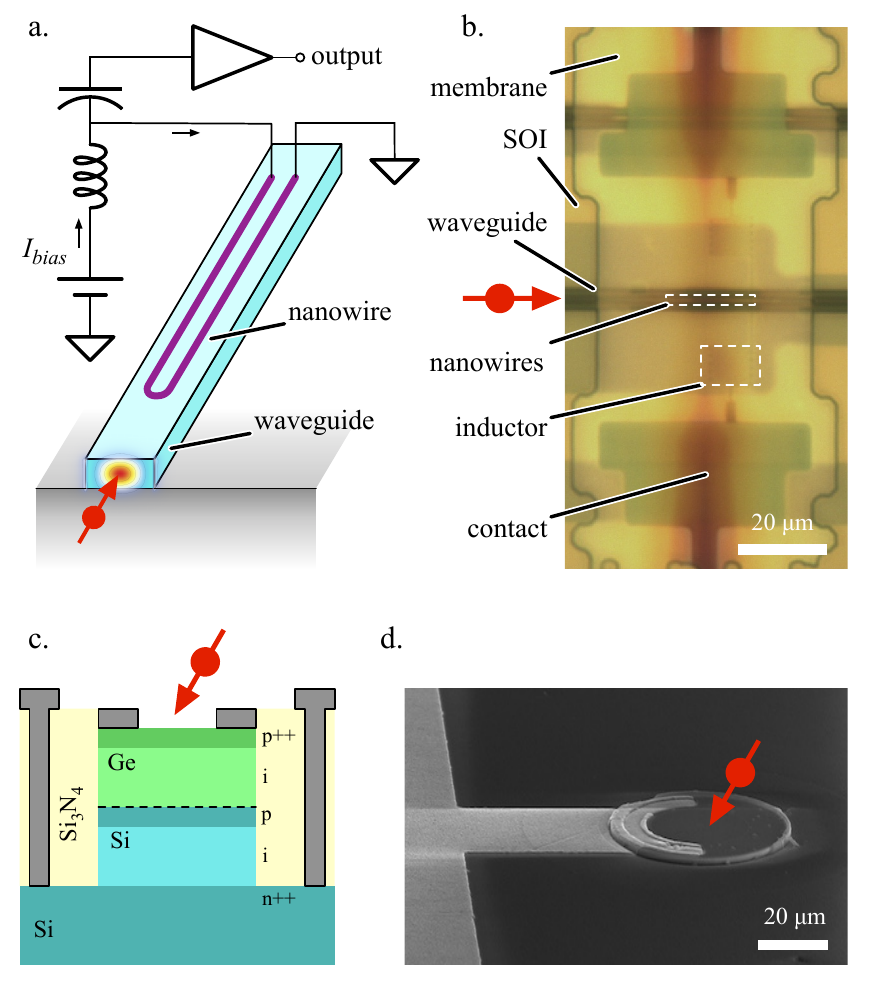}{Optical \caplabel{a} and electron \caplabel{b} micrographs of a technique for improving superconducting nanowire single-photon detector (SNSPD) yield, from reference \cite{Najafi:2015ey}. A batch of SNSPDs is fabricated on a Si$_3$N$_4$ membrane, selected based on performance, and transferred onto a silicon waveguide for use. \caplabel{c} Cross-section view of a vertically-coupled germanium avalanche photodiode from \cite{Warburton:2013hk}, and \caplabel{d} electron micrograph of the fabricated device. Image credit: N. C. Harris, G. S. Buller.}

Quantum opticians have always worked in the wavelength bands least limited by photon sources, optics, and single-photon detectors. Initial experiments sourced correlated photons from atomic `cascades' in sodium \cite{Kimble:1977iw} and calcium \cite{Aspect:1981ga} vapours, which emit visible photons. Visible-wavelength optics were also easily sourced. Later, with the advent of parametric down-conversion photon-pair sources and silicon avalanche photodiode (APD) detectors, experiments shifted to the region of maximum silicon absorption and APD sensitivity, around 800~\nm. Recently, driven by the enormous optical telecommunications industry, activity has been building in the telecoms band around 1.55~\micron. Here, long-range transmission over optical fibre is possible, and the low-cost and high-quality tools of the telecoms industry can be exploited---lasers, (bright-light) detectors, and optics in bulk and fibre. This region is also amenable to waveguides made of silicon, being within silicon's 1100-\nm bandgap. It does present a single-photon detection problem, however, as the community's silicon-APD workhorse is not sensitive in this spectral region.
 
At cryogenic temperatures\footnote{In the range of 1 to 10 K, where non-dilution Gifford-McMahon fridges are relatively cheap and readily available.}, the recently developed superconducting nanowire single-photon detectors (SNSPDs, Fig.~\ref{fig:snspd}a,\cite{Goltsman:2001ea, Dauler:2014bq}) are sensitive at a wide range of wavelengths, including in the 1.55~\micron band.  They realize near-ideal detection characteristics, and require only a single patterning step (typically via electron beam lithography, EBL). So far, a modest exploration of superconducting materials has been carried out. Devices have primarily been based on polycrystalline films of NbN \cite{Goltsman:2001ea, Heath:2015da, Rath:2015um}, but yield considerations have put a new focus on the amorphous superconductors:  WSi \cite{Marsili:2012th, Verma:2014dw, Beyer:2015ga},  MoSi \cite{Verma:2015vg, Korneeva:2014es}, and MoGe \cite{Verma:2014ez}.  Detectors have been patterned atop waveguides of  various materials, mostly using NbN films \cite{Beyer:2015ga, Pernice:2012bc, Sprengers:2011era, Rath:2015um, Reithmaier:2015ju, Reithmaier:2013jk, Najafi:2015ey}.  By building intimate waveguide-nanowire contact into these devices,  very short meanders and near-unit absorption can be achieved simultaneously. Najafi, Mower, Harris \emph{et al.} have demonstrated a different approach to high-yield waveguide-coupled detectors \cite{Najafi:2015ey}. They fabricated several NbN SNSPDs, each on its own Si$_3$N$_4$ membrane, then---after verifying its performance---they transferred it to a silicon waveguide on a different substrate. One such membrane is pictured in Fig.~\ref{fig:snspd}b.

No waveguide-coupled single-photon detector has yet been demonstrated at or near room temperature. There is hope, though, that one day a detector such as the vertically coupled germanium APD demonstrated by Warburton \emph{et al.} \cite{Warburton:2013hk} (shown in Figs. \ref{fig:snspd}c and \ref{fig:snspd}d) could fill the role, with sufficient performance.

\subsection{Optical switches}
\label{sec:switches}

\makefig[b]{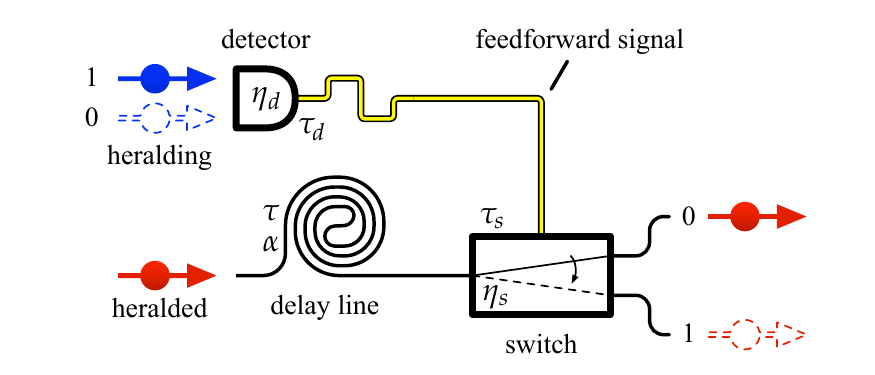}{Conceptual picture of the integration of several functionalities to implement an elementary feedforward operation. The path taken by one single photon (red, `heralded') is controlled by another incoming photon (blue, `heralding'). Upon detection of the heralding photon, a switch is triggered while the heralded photon is stored in a delay line for a time $\tau$. }

LOQC was not taken seriously until 2001, when Knill, Laflamme, and Milburn discovered that by adding extra photons, and using feedforward, one could effectively implement the missing nonlinear interactions required to implement a universal gate set \cite{KLM2001}. In practice a feedforward step requires reconfiguration of a large photonic network in a short time frame. A fast, low loss, low noise, non-blocking optical switch has a central role to play in the LOQC architecture \cite{KLM2001, Li:2015ue}. 

A simple example of feedforward is shown schematically in Fig. \ref{fig:feedforward}. \new{Here, a detector measures a heralding photon (latency $\tau_d$, efficiency $\eta_d$) and actuates an optical switch (settling time $\tau_s$, transmission $\eta_s$), while a heralded photon is delayed by $\tau = \tau_d + \tau_s$ in a delay line with propagation loss and group velocity $\alpha$ and $v_g$, respectively. Delay lines are discussed in detail in section \ref{sec:delays}. The heralded photon is transmitted with efficiency $\eta = \eta_d\eta_s\mathrm{exp}{[\alpha \tau v_g]}$, showing the tradeoff between efficiency and speed in detector, delay, and switch.}
 
Heat- and MEMS-based phase tuners (e.g. \cite{ha-oe-22-10487, wa-oe-38-733} and \cite{se-optica-3-64}) are extremely versatile when it comes to reconfiguring large complex networks. However, they lack the speed (operating at most in the MHz regime) for rapid, time-of-flight feedforward applications, like the one outlined above.
 
Electro-optic (\chitwo) modulators have proven themselves to be strong candidates, with a response time limited only by their driving electronics---they are routinely operated at 40~Gb/s \cite{lithium_niobate_reviewIEEE}. They are also capable of handling quantum states of light, with no additional inherent loss from the electro-optic effect \cite{Predevelclusterstate2007, bo-prl-ln-2012}.
 
Silicon, however, being a centro-symmetric crystal, naturally has $\chitwo = 0$. The usual approaches for implementing switches in silicon rely on modulating carrier concentrations in the waveguides to change the refractive index \cite{th-np-4-518}, offering switching rates in the tens of GHz. Unfortunately, these effects are accompanied by phase-dependent loss, and by noise when operated in a forward-biased configuration \cite{bu-prb-11-5848, SAIEE-103-18}. Nonlinear optics can be leveraged to implement all-optical switching, either relying on optical free-carrier generation \cite{Al-nat-431-1081}, or on the Kerr effect and cross-phase modulation (XPM) \cite{so64374}. These switch architectures, however, suffer an increased complexity arising from the need to add and remove a bright optical pump. A fibre implementation of this scheme was used to switch single photons by Hall \emph{et al.} in 2011 \cite{Hall:2011ki}.

Another approach is to induce a \chitwo in silicon. This can be achieved by breaking silicon's lattice symmetry using strain \cite{Nature2006SiliconKhi2}, or by applying a strong DC electric field. This latter approach has been demonstrated in a reverse-biased PIN modulator \cite{Timurdogan:2016wh}, using standard silicon manufacturing techniques. Alternately, high-\chitwo materials can be directly integrated with silicon. Aluminum nitride has been integrated with silicon \cite{Xiong-AlN:2012}; a silicon-organic hybrid device has shown $>$\,40~Gb/s rates \cite{Al-OptExp-20110606}, and barium titanate has been epitaxially grown on silicon, and used to build ring and Mach-Zehnder modulators \cite{Ab-natcomm-4-1671, Xi-nl-14-1419, Abel:2013fj}.


\subsection{High-Extinction Filters}

\makefig{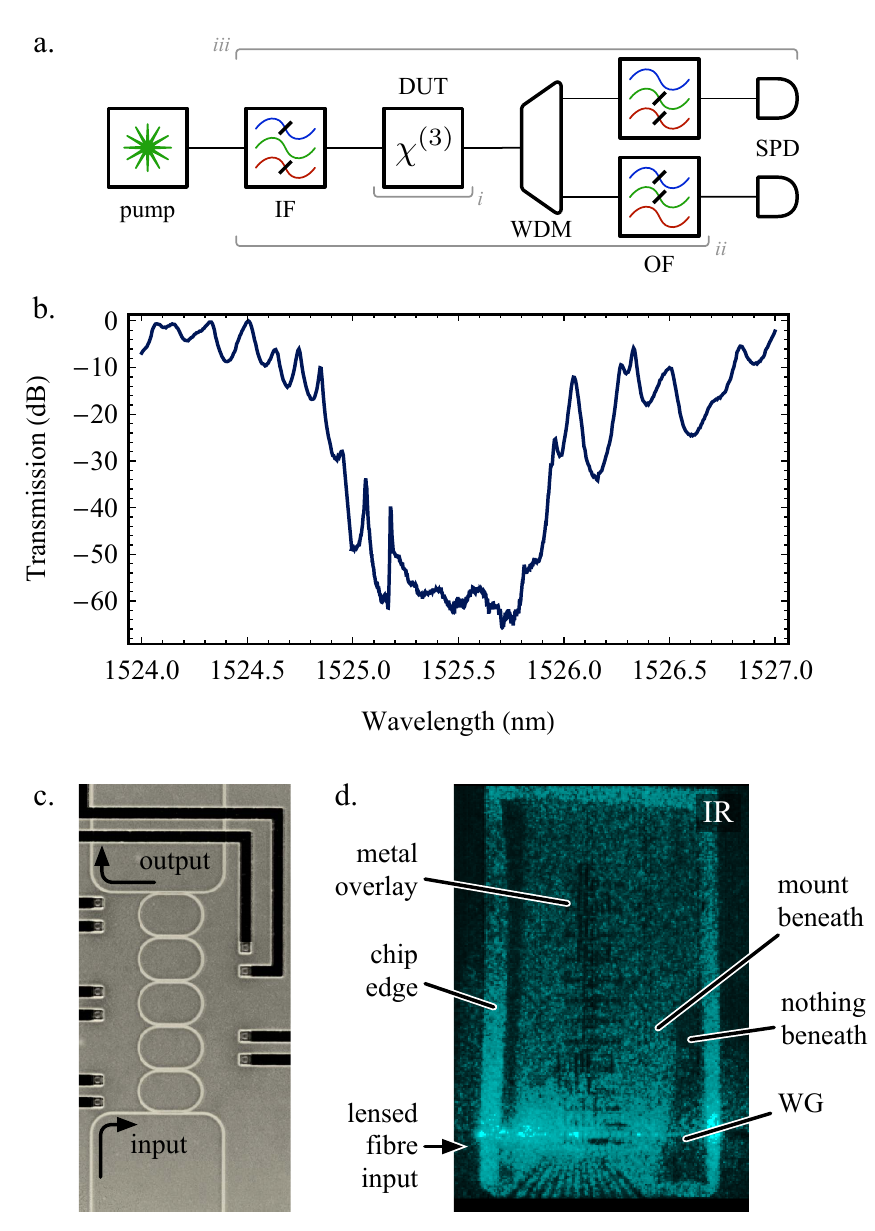}{\new{\caplabel{a} Filter block diagram of a quantum photonic device (DUT) utilizing nonlinear photon-pair sources; input filter (e.g. for removing laser ASE), IF; wavelength-division multiplexer, WDM; output filter, OF; single-photon detectors, SPD. Labels \emph{i--iii} indicate progressive levels of on-chip integration. Typical devices integrate only the source and linear optics (\emph{i}).} \caplabel{b} Transmission spectrum of a single filtering chip, collected between adjacent fibre array channels, from ref. \cite{Harris:2014kj}. \caplabel{c} Optical micrograph of CROW-based filter, from ref. \cite{Ong:2013uh}. \caplabel{d} False-color infrared micrograph of 1.55-\micron scattered laser light inside the substrate of an edge-coupled SOI chip. Image credit: L. Kling.}

In many circumstances, we will be forced to operate photon-pair sources and single-photon detectors in close proximity---in the same subsystem, and likely on the same die. Every effort must be made to ensure that such detectors are not triggered by one of the ca. $10^{10}$ photons ($\sim$\,1~nJ) which compose each of the source's pump pulses. Let us assume that a device performs a quantum information task which cannot withstand more than 1 false detection event in every 100 pulses. To meet this threshold, we need spectral filters which can isolate every single quantum-information-carrying photon from its generating pump pulse. These filters have a highly demanding specification: they must provide at least 120~dB of isolation between the pump and single photons, and---like all quantum optical elements---they should do this with minimal single-photon loss. Fig. \ref{fig:filters}a details the configuration of the various high-extinction filters and wavelength-division multiplexers (WDM) required in a typical device, along with the various stages of progress in the on-chip integration of these components.

Towards these high-extinction filters, several strategies have been employed. Devices have used combinations of coherent and incoherent coupling between sub-structures to achieve these extreme extinction ratios. CROW-based filters (Fig. \ref{fig:filters}c) have been reported to achieve 100~dB extinction \cite{Ong:2013uh}. The combination of corrugated-waveguide Bragg reflectors and ring resonators have enabled a correlation measurement of photon pairs entirely without further external filtering \cite{Harris:2014kj} (the pump transmission spectrum of this device is plotted in \ref{fig:filters}b). A similar result has been recently achieved using cascaded lattice filters \cite{mat-filter2chips-2016}. Finally, the generation and demultiplexing of photon pairs has been demonstrated on a hybrid silicon-silica waveguide platform by Matsuda \emph{et al.}, but this device required further external filtering to fully suppress the experiment's bright pump \cite{Matsuda:2014cy}.

These filtering experiments all have one feature in common: they rely on two physically separated filtering stages, each on its own chip, linked by an optical fibre. This is because the extinction ratio is not limited by a given filter's design or fabrication, but rather by pump leaking \emph{around} that filter in the cladding, fibre-coupling apparatus, and substrate. Fig.~\ref{fig:filters}d shows an infrared micrograph of an edge-coupled SOI chip, highlighting this scattering. So far, the use of two distinct, well-separated chips has been necessary in these proof-of-principle devices to suppress this scattering. Further engineering is required to mitigate this scattering, and to push the boundaries of integration in monolithic systems.

\begin{table*}[!t]
\renewcommand{\arraystretch}{1.3}
\caption{Technology Platforms for Quantum Photonics}
\label{table}
\centering
\newcommand\pt[1][1]{\Repeat{#1}{$\bullet$}} 
\def\tablecolumnwidth{1.5cm}
\newcolumntype{C}{>{\centering\arraybackslash}p{\tablecolumnwidth}}
\begin{tabular}{p{2cm}|CCCCCCC}
Metric & Silicon & Silica & Direct-Write & Si$_3$N$_4$ & InP & GaAs & LiNbO$_3$ \\
\hline
Density \hfill ($1/r^2$)	& \pt[5] & \pt[2] & \pt[1] & \pt[3] & \pt[3] & \pt[5] & \pt[2] \\
Loss \hfill ($1/\alpha r$)	& \pt[5] & \pt[4] & \pt[2] & \pt[4] & \pt[3] & \pt[4] & \pt[3] \\
Passive optics					& \pt[5] & \pt[5] & \pt[4] & \pt[4] & \pt[2] & \pt[1] & \pt[3] \\
Active optics						& \pt[3] & \pt[1] & \pt[1] & \pt[1] & \pt[4] & \pt[3] & \pt[5] \\
Photon sources					& \pt[3] & \pt[1] & \pt[2] & \pt[4] & \pt[2] & \pt[5] & \pt[4] \\
\hline
\end{tabular}
\end{table*}

\subsection{Fibre-to-Chip Coupling}
\label{sec:fibtochip}


Efficient coupling between the nanoscopic silicon waveguide core and a standard optical fibre has been a key challenge since the early days of silicon photonics. Two main methods have emerged. One, edge-coupling, requires a small-mode lensed fibre ($\sim$\,2--5~\micron spot) to couple light into an adiabatic spot-size converter \cite{Almeida:2003tp}. The specialty fibre and small mode size make it impractical couple multiple fibres at the same time using this method. A second method involves fabricating a second-order grating, to vertically couple light directly between a cleaved optical fibre and a waveguide \cite{ro-oe-10-1364,Vivien:06}. This method has an inherently limited bandwidth, due to the grating, but these gratings can be densely placed and accessed with multi-core fibres and v-groove array packages. Both methods have shown tremendous efficiency improvements, with per-channel coupling losses now below 1~dB \cite{Cardenas:2014tp, Notaros:2016vw}. Low-loss interconnectivity could enable quantum key distribution between silicon transceivers, and frees quantum system architectures from strictly monolithic integration.

Both types of couplers are routinely used for injecting pump fields and for collecting single photons. Polarization-splitting two-dimensional grating couplers \cite{Taillaert:2003gc} are an appealing solution for converting path-encoded qubits (stable on a monolithic chip, but not in fibre) to polarisation-encoded qubits (better suited for short fibre and free-space links). Olislager \emph{et al.} reported an entangled photon-pair source using a two-dimensional grating \cite{Olislager:2013gi}, and Wang \emph{et al.} used this structure for chip-to-chip quantum communication \cite{Wang:2016eo}.

\subsection{Delay Lines}
\label{sec:delays}

From time-bin encoded QKD \cite{Harada2008Generation} to one-way quantum computing \cite{ra-prl-22-5188}, many photonic implementations of quantum information tasks require memory. Storing a photon for an arbitrary amount of time, via a quantum memory, is a challenging task. This can be achieved, for example, using nonlinear interactions \cite{Ly-np-3-706, be-delay-icton-2013} but such memories are, for now, difficult to manufacture and integrate, and have a significant control overhead. Plasma dispersion based tunable delay lines have been demonstrated on silicon \cite{kh-oe-19-11780}, but they have inherent losses and their single-photon operation has not yet been tested. Discrete programmable \emph{delay lines} can be made from several different optical structures \cite{melloni-delayline-2010}: CROWs \cite{po-josab-21-1665}, photonic crystals \cite{Takesue:2013db}, or sequences of unbalanced MZIs \cite{xi-oe-22-22707}. The latter approach has been used in a recent QKD demonstration \cite{Sibson:2015uv}.

In many applications---including in a quantum repeater \cite{az-ncomm-6-6787}, QKD \cite{Harada2008Generation}, and one-way quantum computing \cite{GimenoSegovia:2014uf}---a fixed duration memory is enough. Quantum states of light can simply be stored in a long waveguide. Low-loss delay lines (0.024~dB/ns) compatible with the silicon platform have been developed \cite{lee-natcom-3-867, ba-oe-21-544} with demonstrations of delays of up to 130~ns. With fibre-to-chip coupling efficiency continuing to increase (see previous section), low-loss optical fiber may represent the ultimate delay solution. While the footprint of a delay line can be large, this type of memory has inherently high bandwidth, no noise, no requirement for active control, and it works in all environments, from millikelvin to room temperature.

\section{Other Platforms}

Silicon photonics is a strong contender for the quantum photonics crown, but is not without able challengers. In Table~\ref{table}, we qualitatively and quantitatively compare silicon, as a platform for quantum optics, with several other leading technology platforms. This comparison is not exhaustive, and is by necessity subjective; we have tried to make it quantitative wherever possible. We assess the six main platforms used for on-chip quantum optics---silicon, etched silica, direct-write, silicon nitride, indium phosphide, and lithium niobate---in five categories---integration density, loss per bend, passive and active optics, and photon sources. 

Our density and loss metrics are weighted by published bend radii ($r$) and propagation loss ($\alpha$) \cite{Vlasov:2004bo, Ou:2003ci, Crespi:2011cy, Moss:2013kv, Van:2001je, Marpaung:2013gq}. \new{We use bend footprint and loss per bend (proportional to $r^2$ and $\alpha r$, respectively) as indicators of density and total loss.} Our passive optics metric is based on yield and repeatability, whereas our active optics metric depends on modulator bandwidth and loss. We include an estimation of each platform's photon source performance and compatibility. Silicon photon-pair sources based on SFWM have been well-explored, revealing themselves to be versatile but vulnerable to TPA-based effects (see \ref{sec:sources}). Recent results in silicon nitride \cite{Ramelow:2015dx, Ramelow:2015uf} and direct-write silica waveguides \cite{Hiemstra:2015ue, Spring:2016ti, Spring:2013eg}, also based on SFWM, show promise. Gallium arsenide and lithium niobate both host photon-pair sources based on SPDC \cite{Autebert:2015cn, Orieux:2013jv, Horn:2012, Tanzilli:2011hk}; GaAs also supports sources of true single photons, from InAs and InGaAs quantum dots \cite{Reithmaier:2015ju, Reithmaier:2013jk, Claudon:2010cb}. Finally, indium phosphide has the advantage of a mature classical photonics ecosystem, including an electrically pumped laser; for QKD systems requiring just one photon at a time, this laser is an adequate replacement \cite{Sibson:2015uv}. We exclude detectors from this list, as the current dominant technology (the SNSPD) has now been demonstrated with high performance integrated with all the above material systems, with the exception of indium phosphide.

Silicon (unsurprisingly) scores well overall in our comparison. It has natural strengths in density, loss, and passives, and weaknesses in actives and photon sources, due to lack of a high-speed modulator with low-loss, and lack of a high purity and heralding efficiency photon-pair source.

\section{The Future of Quantum Optics in Silicon}

In the remainder of this review, we discuss the challenges facing silicon quantum photonics, and potential avenues of enquiry to overcome them. 

\subsection{Integration}

The main challenge for the future quantum photonic engineer will be to take all the various components outlined in Section~\ref{sec:review}, and combine them on a single die or in a single package. Several obstacles lie before him or her: all components must operate in a common environment, and different components pull this environment in different directions (e.g. cold superconducting detectors vs. hot CMOS electronics); all components must use manufacturing processes which are mutually compatible; and this future system must be packageable, so as to be robust and reliable.


It is increasingly likely that the future of quantum photonics will be cold. There has been no demonstration of a waveguide-coupled detector operating at room-temperature\footnote{The development of a room-temperature detector is not inconceivable, but may prove challenging;  a waveguide-coupled, CMOS-compatible germanium avalanche photodiode \cite{Warburton:2013hk} may be a good candidate.}. SNSPDs operate around 2~K, and offer very high performance. Superconducting detectors present a devil's bargain: they represent a near-perfect single-photon detector, with the caveat that \emph{everything else must work at cryogenic temperatures}. This caveat casts doubt on the thermal tuners used in most demonstrations to date; it also presents challenges for silicon's carrier-based modulators. It requires careful accounting for system energy budgets. Packaging and testing must adapt to an environment of enormous temperature swings. The absence of materials data at low temperatures makes the development of novel materials more difficult. The unforeseen and unforeseeable effects of this transition are legion.

More pragmatically, silicon quantum photonics at low temperatures may hold several advantages. Compact integrated optics can be readily fit into the confines of a cryostat, in contrast with bulk- or fibre-optic systems. Nonlinear optical effects are largely temperature-independent: \chithree effects work nearly as well at low temperature\cite{Sun:2013db}. SFWM photon-pair sources, as well as XPM and FWM, will continue to work. Thermal-expansion-based straining, used to produce \chitwo in silicon\cite{Jacobsen:2006up}, could benefit from the larger deposition-to-operation temperature difference; \chitwo and \chithree  electro-optic modulators could be invaluable for low-temperature switching. Finally, the effort expended to make silicon photonics \emph{athermal} would be unnecessary: silicon's low-temperature thermo-optic coefficient is $10\,000$ times smaller than its room-temperature value\cite{Sun:2013db, Komma:2012dj}.
We can be optimistic that silicon photonics can accommodate either front- or back-end integrated electronics, for implementing the feedforward control required (Fig.~\ref{fig:feedforward}). CMOS electronics can be made to work at cryogenic temperatures \cite{Ekanayake:2010fj, NAGATA:2011hj, Anonymous:1974tr}, and superconducting electronics \cite{Likharev:1991dg, Yamashita:2012ge} could serve a naturally complementary role.

Cold silicon quantum photonics certainly presents challenges, but it comes with benefits too. The biggest prize of all, though, is access---enabled by \emph{integrated} single-photon detection---to previously unimaginable system architectures. The multiplexing of single photons \cite{Migdall:2002} is only the first of these architectural shifts.


Several components that we have discussed require materials and methods which are well beyond those present in a typical CMOS foundry. SNSPDs call for sputtered layers of exotic metals; fast, low-loss modulators may require exotic materials like barium titanate \cite{Ab-natcomm-4-1671} or nonlinear polymers \cite{Al-OptExp-20110606}; and grating fibre-to-chip couplers benefit from engineered substrates \cite{Kuno:2015by, Zhang:2014hn}. All these materials and processes have been integrated with silicon photonics, but inevitably there will be challenges in integrating them with each other.

\subsection{Two-Photon Absorption}

Two-photon absorption dominates discussions of nonlinear optics in silicon. TPA is occasionally useful (e.g.\cite{Pelc:2014hz}), but is more often problematic. When we want to use the {Kerr effect} directly---via SPM, XPM, or FWM---the presence of nonlinear loss fundamentally reduces efficiency. In the case of SFWM, this translates to a fundamental reduction in \emph{heralding} efficiency, which will become intolerable as time goes on. Furthermore, when two photons are absorbed, they excite an electron into the conduction band. This electron then adds further loss, via free-carrier absorption, and modifies the refractive index, via free-carrier dispersion (FCD) and by the thermo-optic effect, when it eventually relaxes. In addition to improving heralding, banishing TPA will yield several corollary benefits in techniques which will become feasible: XPM all-optical switches \cite{Rambo:ZC8H7tI3, Pelc:2014hz}; FWM frequency-conversion \cite{Turner:2008bc}; Raman lasing \cite{Rong:2005bb}; and FWM and Raman parametric amplification \cite{Liu:2010by, Espinola:2004kv}.

\makefig{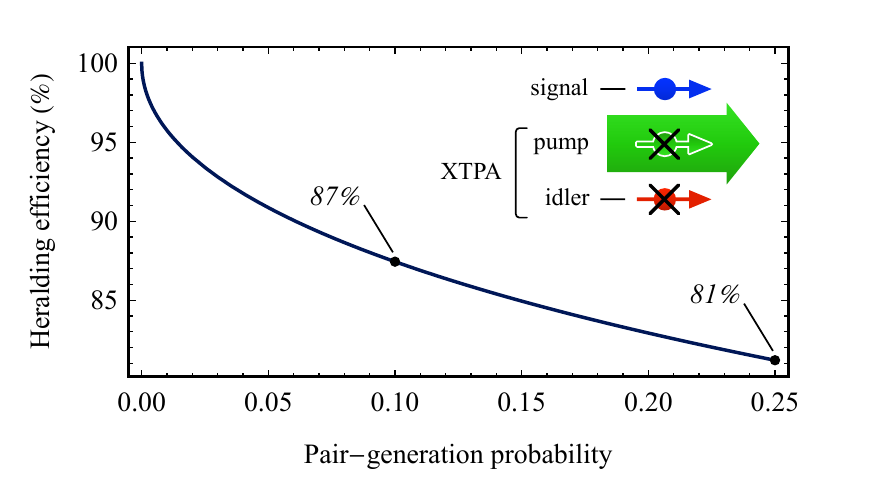}{Maximum heralding efficiency for phase-matched SFWM pair generation in silicon (Eq. \ref{eq:nlloss2}), with $n_2 = 6.3\times 10^{-18} \ \mathrm{m}^2\cdot\mathrm{W}^{-1}$ and $\alpha_2 = 6.14\times 10^{-12} \ \mathrm{m}\cdot\mathrm{W}^{-1}$, at a wavelength of 1.55~\micron. A schematic view of XTPA in a SFWM source is shown in the inset.}

\subsubsection{Effect on heralding efficiency}

As pointed out by Husko et al. in \cite{Husko:2013bh}, XTPA decreases the heralding efficiency of photon-pair sources in silicon. Assuming phase-matched SFWM with weak squeezing, the extra uncorrelated loss on the signal channel, relative to the idler channel, for a given pair-generation probability $p$, is
\begin{equation}
	\eta_\mathrm{XTPA} \approx \frac{1}{(1 + \frac{\alpha_2}{k_0 n_2}\sqrt{\frac{p}{\Delta t \Delta \nu_c}})^2},
	\label{eq:nlloss2}
\end{equation}
where $\alpha_2$, $n_2$, $k_0$, $\Delta t$, and $\Delta \nu_c$ are respectively: the TPA coefficient, nonlinear refractive index, vacuum wavenumber, pump pulse time, and collection bandwidth. This extra loss, plotted in Fig. \ref{fig:xtpaheralding}, is intrinsic to SFWM in silicon at 1.55~\micron, and represents a limit on the attainable heralding efficiency.

Solutions to the TPA problem fall into two broad categories: a switch to a photon-pair source material which exhibits no or greatly reduced TPA; or a switch to a longer wavelength, beyond silicon's two-photon band edge. We discuss these two categories below.

\subsubsection{Larger bandgap materials} 
Large \chithree non-linearity can be obtained from organic materials and can be leveraged using slot waveguides \cite{Leuthold:2009gq}. Amorphous silicon is also a promising candidate \cite{Pelc:2014hz}. However Raman scattering over a large bandwidth will accompany pair generation in these amorphous materials which will have to be operated accordingly\footnote{\new{Low-temperature operation may help suppress Raman scattering.}}.
The addition of \chitwo materials such as GaN \cite{Xiong:11} or AlN \cite{Xiong-AlN:2012} may also provide the option of using SPDC instead of SFWM to generate photon pairs.

\subsubsection{Longer wavelengths}

A shift to longer wavelengths, nearer silicon's two-photon bandgap (2.2~\micron), may be the simplest approach. 2-\micron SOI devices maintain a high nonlinearity, while almost completely eliminating nonlinear absorption \cite{Wang:2013jg}, allowing silicon to keep its nonlinear-optical crown\cite{Jalali:2010fo}. Complexity is shifted entirely to the optics: new pump sources, and detectors must be developed to operate at these longer wavelengths. What makes this approach credible are recent developments on both fronts: high-powered thulium- and holmium-doped fibre lasers are becoming available, and measurements on the amorphous superconductors are showing sensitivity to long wavelengths\cite{Marsili:2013hy, Baek:2011cs, Korneeva:2011de}. Indeed, SNSPDs were first developed to detect millimetre-wave signals\cite{Gershenzon:1990mm}. Provided these optical challenges can be met, the long-wavelength route could be a promising one.

%


\section{Conclusion}

Quantum photonic applications are beginning to come into technological reach, as evidenced by the recent demonstration of a fully chip-based QKD system \cite{Sibson:2015uv}, relying on an indium phosphide transmitter and a silicon oxynitride receiver. Commercializers of this technology could rapidly find that, for mass-deployment, silicon photonics is the only route to the associated mass-manufacture.


Multi-project wafer (MPW) services allow many small research teams to share the cost of state-of-the-art silicon manufacturing for small-batch and prototype devices\footnote{Reference \cite{Chrostowski:2015bk} contains a recent list of silicon photonics MPW services.}. These services give the research community the ability to test and prove small-scale quantum photonic designs without needing the resources to fabricate or commission entire wafers. In the near and medium terms, these services could seriously accelerate the development of silicon quantum photonics, allowing the community to incrementally address the remaining issues, and to test strategies for integration and packaging without large-scale investment.

Our rising quantum photonic technological capability, and the falling thresholds for linear optical quantum computation (e.g. \cite{GimenoSegovia:2014uf, Li:2015ue, Barrett:2010goa}), are poised to meet in the middle. At its core, silicon technology provides the only viable route to assembling systems of millions of components. So, in addition to its natural technical advantages, it may represent the only option for constructing quantum computers which run on light. In this review, we have provided an overview of the technological modules required by future quantum photonic systems, and we have elaborated on the strengths, weaknesses, context, and future challenges for silicon as a springboard for photonic quantum technologies.

\section*{Acknowledgment}
This work has been supported by the Engineering and Physical Science Research Council UK (EPSRC), the European Research Council (ERC), the FP7 European project BBOI, and the US Army Research Office (ARO). J. W. S. acknowledges support from the Natural Sciences and Engineering Research Council of Canada (NSERC). M. G. T. acknowledges support from an EPSRC Early Career Fellowship, and from an ERC starting grant. 
We are grateful to N. C. Harris and G. S. Buller for providing material for Figs.~\ref{fig:mitpnp}, \ref{fig:snspd}, and \ref{fig:filters}.

\ifCLASSOPTIONcaptionsoff
  \newpage
\fi



%

\bibliography{bibli,bibjws,BibliographyUTF8} 

\begin{thebibliography}{100}
\expandafter\ifx\csname url\endcsname\relax
  \def\url#1{\texttt{#1}}\fi
\expandafter\ifx\csname urlprefix\endcsname\relax\def\urlprefix{URL }\fi
\providecommand{\bibinfo}[2]{#2}
\providecommand{\eprint}[2][]{\url{#2}}

\bibitem{Bennett:1992jx}
\bibinfo{author}{Bennett, C.~H.}, \bibinfo{author}{Brassard, G.} \&
  \bibinfo{author}{Mermin, N.~D.}
\newblock \bibinfo{title}{{Quantum cryptography without Bell{\textquoteright}s
  theorem}}.
\newblock \emph{\bibinfo{journal}{Phys. Rev. Lett.}}
  \textbf{\bibinfo{volume}{68}}, \bibinfo{pages}{557--559}
  (\bibinfo{year}{1992}).

\bibitem{Ekert:1991kl}
\bibinfo{author}{Ekert, A.~K.}
\newblock \bibinfo{title}{{Quantum cryptography based on Bell{\textquoteright}s
  theorem}}.
\newblock \emph{\bibinfo{journal}{Phys. Rev. Lett.}}
  \textbf{\bibinfo{volume}{67}}, \bibinfo{pages}{661--663}
  (\bibinfo{year}{1991}).

\bibitem{Ladd:2010kq}
\bibinfo{author}{Ladd, T.~D.} \emph{et~al.}
\newblock \bibinfo{title}{{Quantum computers.}}
\newblock \emph{\bibinfo{journal}{Nature}} \textbf{\bibinfo{volume}{464}},
  \bibinfo{pages}{45--53} (\bibinfo{year}{2010}).

\bibitem{AspuruGuzik:2012ho}
\bibinfo{author}{Walther, P.}
\newblock \bibinfo{title}{{Photonic quantum simulators}}.
\newblock \emph{\bibinfo{journal}{Nature Physics}}
  \textbf{\bibinfo{volume}{8}}, \bibinfo{pages}{285--291}
  (\bibinfo{year}{2012}).

\bibitem{Giovannetti:2011jka}
\bibinfo{author}{Giovannetti, V.}, \bibinfo{author}{Lloyd, S.} \&
  \bibinfo{author}{Maccone, L.}
\newblock \bibinfo{title}{{Advances in quantum metrology}}.
\newblock \emph{\bibinfo{journal}{Nature Photonics}}
  \textbf{\bibinfo{volume}{5}}, \bibinfo{pages}{222--229}
  (\bibinfo{year}{2011}).

\bibitem{OBrien:2009eu}
\bibinfo{author}{O'Brien, J.~L.}, \bibinfo{author}{Furusawa, A.} \&
  \bibinfo{author}{Vuckovic, J.}
\newblock \bibinfo{title}{{Photonic quantum technologies}}.
\newblock \emph{\bibinfo{journal}{Nature Photonics}}
  \textbf{\bibinfo{volume}{3}}, \bibinfo{pages}{687--695}
  (\bibinfo{year}{2009}).

\bibitem{Politi:2008jg}
\bibinfo{author}{Cryan, M.~J.}, \bibinfo{author}{Rarity, J.~G.},
  \bibinfo{author}{Yu, S.} \& \bibinfo{author}{O'Brien, J.~L.}
\newblock \bibinfo{title}{{Silica-on-silicon waveguide quantum circuits}}.
\newblock \emph{\bibinfo{journal}{Science}} \textbf{\bibinfo{volume}{320}},
  \bibinfo{pages}{646--649} (\bibinfo{year}{2008}).

\bibitem{Peruzzo:2010tq}
\bibinfo{author}{Peruzzo, A.} \emph{et~al.}
\newblock \bibinfo{title}{{Quantum walks of correlated photons}}.
\newblock \emph{\bibinfo{journal}{Science}} \textbf{\bibinfo{volume}{329}},
  \bibinfo{pages}{1500--1503} (\bibinfo{year}{2010}).

\bibitem{Shadbolt:2012bw}
\bibinfo{author}{Shadbolt, P.~J.} \emph{et~al.}
\newblock \bibinfo{title}{{Generating, manipulating and measuring entanglement
  and mixture with a reconfigurable photonic circuit}}.
\newblock \emph{\bibinfo{journal}{Nature Photonics}}
  \textbf{\bibinfo{volume}{6}}, \bibinfo{pages}{45--49} (\bibinfo{year}{2012}).

\bibitem{Spring:2013do}
\bibinfo{author}{Spring, J.~B.} \emph{et~al.}
\newblock \bibinfo{title}{{Boson sampling on a photonic chip.}}
\newblock \emph{\bibinfo{journal}{Science}} \textbf{\bibinfo{volume}{339}},
  \bibinfo{pages}{798--801} (\bibinfo{year}{2013}).

\bibitem{Crespi:2013fu}
\bibinfo{author}{Crespi, A.} \emph{et~al.}
\newblock \bibinfo{title}{{Integrated multimode interferometers with arbitrary
  designs for photonic boson sampling}}.
\newblock \emph{\bibinfo{journal}{Nature Photonics}}
  \textbf{\bibinfo{volume}{7}}, \bibinfo{pages}{545--549}
  (\bibinfo{year}{2013}).

\bibitem{Tillmann:2013jv}
\bibinfo{author}{Tillmann, M.} \emph{et~al.}
\newblock \bibinfo{title}{{Experimental boson sampling}}.
\newblock \emph{\bibinfo{journal}{Nature Photonics}}
  \textbf{\bibinfo{volume}{7}}, \bibinfo{pages}{540--544}
  (\bibinfo{year}{2013}).

\bibitem{Sun:2013kl}
\bibinfo{author}{Sun, J.}, \bibinfo{author}{Timurdogan, E.},
  \bibinfo{author}{Yaacobi, A.}, \bibinfo{author}{Hosseini, E.~S.} \&
  \bibinfo{author}{Watts, M.~R.}
\newblock \bibinfo{title}{{Large-scale nanophotonic phased array.}}
\newblock \emph{\bibinfo{journal}{Nature}} \textbf{\bibinfo{volume}{493}},
  \bibinfo{pages}{195--199} (\bibinfo{year}{2013}).

\bibitem{Han:2015jh}
\bibinfo{author}{Han, S.} \emph{et~al.}
\newblock \bibinfo{title}{{Large-scale silicon photonic switches with movable
  directional couplers}}.
\newblock \emph{\bibinfo{journal}{Optica}} \textbf{\bibinfo{volume}{2}},
  \bibinfo{pages}{370--375} (\bibinfo{year}{2015}).

\bibitem{Sun:2015gg}
\bibinfo{author}{Sun, C.} \emph{et~al.}
\newblock \bibinfo{title}{{Single-chip microprocessor that communicates
  directly using light}}.
\newblock \emph{\bibinfo{journal}{Nature}} \textbf{\bibinfo{volume}{528}},
  \bibinfo{pages}{534--538} (\bibinfo{year}{2015}).

\bibitem{Morichetti:2010hp}
\bibinfo{author}{Morichetti, F.} \emph{et~al.}
\newblock \bibinfo{title}{{Roughness Induced Backscattering in Optical Silicon
  Waveguides}}.
\newblock \emph{\bibinfo{journal}{Phys. Rev. Lett.}}
  \textbf{\bibinfo{volume}{104}}, \bibinfo{pages}{033902}
  (\bibinfo{year}{2010}).

\bibitem{Lee:2001kc}
\bibinfo{author}{Lee, K.~K.}, \bibinfo{author}{Lim, D.~R.},
  \bibinfo{author}{Lim, D.~R.}, \bibinfo{author}{Shin, J.} \&
  \bibinfo{author}{Cerrina, F.}
\newblock \bibinfo{title}{{Fabrication of ultralow-loss Si/SiO$_2$ waveguides
  by roughness reduction}}.
\newblock \emph{\bibinfo{journal}{Opt. Lett.}} \textbf{\bibinfo{volume}{26}},
  \bibinfo{pages}{1888--1890} (\bibinfo{year}{2001}).

\bibitem{gr-njp-2004}
\bibinfo{author}{Grangier, P.}, \bibinfo{author}{Sanders, B.} \&
  \bibinfo{author}{Vuckovic, J.}
\newblock \bibinfo{title}{Focus on single photons on demand}.
\newblock \emph{\bibinfo{journal}{New Journal of Physics}}
  \textbf{\bibinfo{volume}{6}} (\bibinfo{year}{2004}).

\bibitem{sc-rmp-81-1301}
\bibinfo{author}{Scarani, V.} \emph{et~al.}
\newblock \bibinfo{title}{{The security of practical quantum key
  distribution}}.
\newblock \emph{\bibinfo{journal}{Reviews of Modern Physics}}
  \textbf{\bibinfo{volume}{81}}, \bibinfo{pages}{1301--1350}
  (\bibinfo{year}{2009}).

\bibitem{Sibson:2015uv}
\bibinfo{author}{Sibson, P.} \emph{et~al.}
\newblock \bibinfo{title}{{Chip-based Quantum Key Distribution}}.
\newblock \emph{\bibinfo{journal}{arXiv}} \bibinfo{pages}{1509.00768}
  (\bibinfo{year}{2015}).

\bibitem{Ei-aip-82-071101}
\bibinfo{author}{Eisaman, M.~D.}, \bibinfo{author}{Fan, J.},
  \bibinfo{author}{Migdall, A.} \& \bibinfo{author}{Polyakov, S.~V.}
\newblock \bibinfo{title}{{Invited Review Article: Single-photon sources and
  detectors}}.
\newblock \emph{\bibinfo{journal}{Review of Scientific Instruments}}
  \textbf{\bibinfo{volume}{82}}, \bibinfo{pages}{071101+}
  (\bibinfo{year}{2011}).

\bibitem{Reithmaier:2015ju}
\bibinfo{author}{Reithmaier, G.} \emph{et~al.}
\newblock \bibinfo{title}{{On-Chip Generation, Routing, and Detection of
  Resonance Fluorescence}}.
\newblock \emph{\bibinfo{journal}{Nano Lett.}} \textbf{\bibinfo{volume}{15}},
  \bibinfo{pages}{5208--5213} (\bibinfo{year}{2015}).

\bibitem{Wei:2014dg}
\bibinfo{author}{Wei, Y.-J.} \emph{et~al.}
\newblock \bibinfo{title}{{Deterministic and Robust Generation of Single
  Photons from a Single Quantum Dot with {99.5\%} Indistinguishability Using
  Adiabatic Rapid Passage}}.
\newblock \emph{\bibinfo{journal}{Nano Lett.}} \textbf{\bibinfo{volume}{14}},
  \bibinfo{pages}{6515--6519} (\bibinfo{year}{2014}).

\bibitem{xi-apl-98-51101}
\bibinfo{author}{Xiong, C.} \emph{et~al.}
\newblock \bibinfo{title}{{Generation of correlated photon pairs in a
  chalcogenide As$_2$S$_3$ waveguide}}.
\newblock \emph{\bibinfo{journal}{Applied Physics Letters}}
  \textbf{\bibinfo{volume}{98}}, \bibinfo{pages}{051101+}
  (\bibinfo{year}{2011}).

\bibitem{wo-pra-80-053815}
\bibinfo{author}{Mauerer, W.}, \bibinfo{author}{Avenhaus, M.},
  \bibinfo{author}{Helwig, W.} \& \bibinfo{author}{Silberhorn, C.}
\newblock \bibinfo{title}{{How colors influence numbers: Photon statistics of
  parametric down-conversion}}.
\newblock \emph{\bibinfo{journal}{Physical Review A}}
  \textbf{\bibinfo{volume}{80}} (\bibinfo{year}{2009}).

\bibitem{Tapster:1998cv}
\bibinfo{author}{Tapster, P.~R.} \& \bibinfo{author}{Rarity, J.~G.}
\newblock \bibinfo{title}{{Photon statistics of pulsed parametric light}}.
\newblock \emph{\bibinfo{journal}{Journal of Modern Optics}}
  \textbf{\bibinfo{volume}{45}}, \bibinfo{pages}{595--604}
  (\bibinfo{year}{1998}).

\bibitem{Gerry:2005hc}
\bibinfo{author}{Gerry, C.~C.} \& \bibinfo{author}{Knight, P.~L.}
\newblock \emph{\bibinfo{title}{{Introductory Quantum Optics.}}}
  (\bibinfo{publisher}{Cambridge University Press},
  \bibinfo{address}{Cambridge, UK}, \bibinfo{year}{2005}),
  \bibinfo{edition}{first} edn.

\bibitem{Helt:2012jj}
\bibinfo{author}{Helt, L.~G.}, \bibinfo{author}{Liscidini, M.} \&
  \bibinfo{author}{Sipe, J.~E.}
\newblock \bibinfo{title}{{How does it scale? Comparing quantum and classical
  nonlinear optical processes in integrated devices}}.
\newblock \emph{\bibinfo{journal}{J. Opt. Soc. Am. B, JOSAB}}
  \textbf{\bibinfo{volume}{29}}, \bibinfo{pages}{2199--2212}
  (\bibinfo{year}{2012}).

\bibitem{Migdall-pra-66-053805}
\bibinfo{author}{Migdall, A.~L.}, \bibinfo{author}{Branning, D.} \&
  \bibinfo{author}{Castelletto, S.}
\newblock \bibinfo{title}{{Tailoring single-photon and multiphoton
  probabilities of a single-photon on-demand source}}.
\newblock \emph{\bibinfo{journal}{Physical Review A}}
  \textbf{\bibinfo{volume}{66}}, \bibinfo{pages}{053805+}
  (\bibinfo{year}{2002}).

\bibitem{Engin:2013fb}
\bibinfo{author}{Engin, E.} \emph{et~al.}
\newblock \bibinfo{title}{{Photon pair generation in a silicon micro-ring
  resonator with reverse bias enhancement}}.
\newblock \emph{\bibinfo{journal}{Op. Ex.}} \textbf{\bibinfo{volume}{21}},
  \bibinfo{pages}{27826--27834} (\bibinfo{year}{2013}).

\bibitem{ag-ol-31-3140}
\bibinfo{author}{Lin, Q.} \& \bibinfo{author}{Agrawal, G.~P.}
\newblock \bibinfo{title}{{Silicon waveguides for creating quantum-correlated
  photon pairs}}.
\newblock \emph{\bibinfo{journal}{Optics Letters}}
  \textbf{\bibinfo{volume}{31}}, \bibinfo{pages}{3140--3142}
  (\bibinfo{year}{2006}).

\bibitem{Husko:2013bh}
\bibinfo{author}{Husko, C.~A.}, \bibinfo{author}{Clark, A.~S.},
  \bibinfo{author}{De~Rossi, A.}, \bibinfo{author}{Combrie, S.} \&
  \bibinfo{author}{Rey, I.~H.}
\newblock \bibinfo{title}{{Multi-photon absorption limits to heralded single
  photon sources}}.
\newblock \emph{\bibinfo{journal}{Sci. Rep.}} \textbf{\bibinfo{volume}{3}},
  \bibinfo{pages}{3087} (\bibinfo{year}{2013}).

\bibitem{Rukhlenko:2012hv}
\bibinfo{author}{Rukhlenko, I.~D.} \& \bibinfo{author}{Agrawal, G.~P.}
\newblock \bibinfo{title}{{Effective mode area and its optimization in
  silicon-nanocrystal waveguides.}}
\newblock \emph{\bibinfo{journal}{Opt. Lett.}} \textbf{\bibinfo{volume}{37}},
  \bibinfo{pages}{2295--2297} (\bibinfo{year}{2012}).

\bibitem{lgh-ol-35-3006}
\bibinfo{author}{Helt, L.~G.}, \bibinfo{author}{Yang, Z.},
  \bibinfo{author}{Liscidini, M.} \& \bibinfo{author}{Sipe, J.~E.}
\newblock \bibinfo{title}{{Spontaneous four-wave mixing in microring
  resonators}}.
\newblock \emph{\bibinfo{journal}{Opt. Lett.}} \textbf{\bibinfo{volume}{35}},
  \bibinfo{pages}{3006--3008} (\bibinfo{year}{2010}).

\bibitem{Silverstone:2015cl}
\bibinfo{author}{Silverstone, J.~W.} \emph{et~al.}
\newblock \bibinfo{title}{{Qubit entanglement between ring-resonator
  photon-pair sources on a silicon chip}}.
\newblock \emph{\bibinfo{journal}{Nature Communications}}
  \textbf{\bibinfo{volume}{6}}, \bibinfo{pages}{1--7} (\bibinfo{year}{2015}).

\bibitem{Kumar:2014gb}
\bibinfo{author}{Kumar, R.}, \bibinfo{author}{Ong, J.~R.},
  \bibinfo{author}{Savanier, M.} \& \bibinfo{author}{Mookherjea, S.}
\newblock \bibinfo{title}{{Controlling the spectrum of photons generated on a
  silicon nanophotonic chip.}}
\newblock \emph{\bibinfo{journal}{Nature Communications}}
  \textbf{\bibinfo{volume}{5}}, \bibinfo{pages}{5489} (\bibinfo{year}{2014}).

\bibitem{Lu:2016wu}
\bibinfo{author}{Lu, X.}, \bibinfo{author}{Jiang, W.~C.},
  \bibinfo{author}{Zhang, J.} \& \bibinfo{author}{Lin, Q.}
\newblock \bibinfo{title}{{Biphoton statistic of quantum light generated on a
  silicon chip}}.
\newblock \emph{\bibinfo{journal}{arXiv}} \bibinfo{pages}{1602.08057}
  (\bibinfo{year}{2016}).

\bibitem{Reimer:2015bv}
\bibinfo{author}{Reimer, C.} \emph{et~al.}
\newblock \bibinfo{title}{{Cross-polarized photon-pair generation and
  bi-chromatically pumped optical parametric oscillation on a chip}}.
\newblock \emph{\bibinfo{journal}{Nature Communications}}
  \textbf{\bibinfo{volume}{6}}, \bibinfo{pages}{8236} (\bibinfo{year}{2015}).

\bibitem{Ramelow:2015uf}
\bibinfo{author}{Ramelow, S.}, \bibinfo{author}{Farsi, A.},
  \bibinfo{author}{Clemmen, S.} \& \bibinfo{author}{Orquiza, D.}
\newblock \bibinfo{title}{{Silicon-Nitride Platform for Narrowband Entangled
  Photon Generation}}.
\newblock \emph{\bibinfo{journal}{arXiv}} \bibinfo{pages}{1508.04358}
  (\bibinfo{year}{2015}).

\bibitem{Sharping:2006tv}
\bibinfo{author}{Sharping, J.~E.} \emph{et~al.}
\newblock \bibinfo{title}{{Generation of correlated photons in nanoscale
  silicon waveguides.}}
\newblock \emph{\bibinfo{journal}{Op. Ex.}} \textbf{\bibinfo{volume}{14}},
  \bibinfo{pages}{12388--12393} (\bibinfo{year}{2006}).

\bibitem{Harada:2011cw}
\bibinfo{author}{Harada, K.-i.} \emph{et~al.}
\newblock \bibinfo{title}{{Indistinguishable photon pair generation using two
  independent silicon wire waveguides}}.
\newblock \emph{\bibinfo{journal}{New J. Phys.}} \textbf{\bibinfo{volume}{13}},
  \bibinfo{pages}{065005} (\bibinfo{year}{2011}).

\bibitem{Harada:2008iy}
\bibinfo{author}{Harada, K.-i.} \emph{et~al.}
\newblock \bibinfo{title}{{Generation of high-purity entangled photon pairs
  using silicon wire waveguide}}.
\newblock \emph{\bibinfo{journal}{Op. Ex.}} \textbf{\bibinfo{volume}{16}},
  \bibinfo{pages}{20368--20373} (\bibinfo{year}{2008}).

\bibitem{Clemmen:2009tc}
\bibinfo{author}{Clemmen, S.} \emph{et~al.}
\newblock \bibinfo{title}{{Continuous wave photon pair generation in
  silicon-on-insulator waveguides and ring resonators}}.
\newblock \emph{\bibinfo{journal}{Op. Ex.}} \textbf{\bibinfo{volume}{17}},
  \bibinfo{pages}{16558--16570} (\bibinfo{year}{2009}).

\bibitem{Matsuda:2012dm}
\bibinfo{author}{Matsuda, N.} \emph{et~al.}
\newblock \bibinfo{title}{{A monolithically integrated polarization entangled
  photon pair source on a silicon chip.}}
\newblock \emph{\bibinfo{journal}{Sci. Rep.}} \textbf{\bibinfo{volume}{2}},
  \bibinfo{pages}{817} (\bibinfo{year}{2012}).

\bibitem{Silverstone:2014fu}
\bibinfo{author}{Silverstone, J.~W.} \emph{et~al.}
\newblock \bibinfo{title}{{On-chip quantum interference between silicon
  photon-pair sources}}.
\newblock \emph{\bibinfo{journal}{Nature Photonics}}
  \textbf{\bibinfo{volume}{8}}, \bibinfo{pages}{104--108}
  (\bibinfo{year}{2014}).

\bibitem{Olislager:2013gi}
\bibinfo{author}{Olislager, L.} \emph{et~al.}
\newblock \bibinfo{title}{{Silicon-on-insulator integrated source of
  polarization-entangled photons}}.
\newblock \emph{\bibinfo{journal}{Opt. Lett.}} \textbf{\bibinfo{volume}{38}},
  \bibinfo{pages}{1960--1962} (\bibinfo{year}{2013}).

\bibitem{Azzini:2012io}
\bibinfo{author}{Azzini, S.} \emph{et~al.}
\newblock \bibinfo{title}{{Ultra-low power generation of twin photons in a
  compact silicon ring resonator}}.
\newblock \emph{\bibinfo{journal}{Op. Ex.}} \textbf{\bibinfo{volume}{20}},
  \bibinfo{pages}{23100--23107} (\bibinfo{year}{2012}).

\bibitem{Grassani:2015ft}
\bibinfo{author}{Grassani, D.} \emph{et~al.}
\newblock \bibinfo{title}{{Micrometer-scale integrated silicon source of
  time-energy entangled photons}}.
\newblock \emph{\bibinfo{journal}{Optica}} \textbf{\bibinfo{volume}{2}},
  \bibinfo{pages}{88} (\bibinfo{year}{2015}).

\bibitem{Wakabayashi:2015cl}
\bibinfo{author}{Wakabayashi, R.} \emph{et~al.}
\newblock \bibinfo{title}{{Time-bin entangled photon pair generation from Si
  micro-ring resonator}}.
\newblock \emph{\bibinfo{journal}{Op. Ex.}} \textbf{\bibinfo{volume}{23}},
  \bibinfo{pages}{1103} (\bibinfo{year}{2015}).

\bibitem{Preble:2015tb}
\bibinfo{author}{Preble, S.~F.} \emph{et~al.}
\newblock \bibinfo{title}{{On-Chip Quantum Interference from a Single Silicon
  Ring Resonator Source}}.
\newblock \emph{\bibinfo{journal}{arXiv}} \bibinfo{pages}{1504.04335}
  (\bibinfo{year}{2015}).

\bibitem{Harris:2014kj}
\bibinfo{author}{Harris, N.~C.} \emph{et~al.}
\newblock \bibinfo{title}{{Integrated Source of Spectrally Filtered Correlated
  Photons for Large-Scale Quantum Photonic Systems}}.
\newblock \emph{\bibinfo{journal}{Phys. Rev. X}} \textbf{\bibinfo{volume}{4}},
  \bibinfo{pages}{041047} (\bibinfo{year}{2014}).

\bibitem{sa-apl-107-131101}
\bibinfo{author}{Savanier, M.}, \bibinfo{author}{Kumar, R.} \&
  \bibinfo{author}{Mookherjea, S.}
\newblock \bibinfo{title}{{Optimizing photon-pair generation electronically
  using a p-i-n diode incorporated in a silicon microring resonator}}.
\newblock \emph{\bibinfo{journal}{Applied Physics Letters}}
  \textbf{\bibinfo{volume}{107}}, \bibinfo{pages}{131101+}
  (\bibinfo{year}{2015}).

\bibitem{Reimer:2014ev}
\bibinfo{author}{Reimer, C.} \emph{et~al.}
\newblock \bibinfo{title}{{Integrated frequency comb source of heralded single
  photons}}.
\newblock \emph{\bibinfo{journal}{Op. Ex.}} \textbf{\bibinfo{volume}{22}},
  \bibinfo{pages}{6535--6546} (\bibinfo{year}{2014}).

\bibitem{He:2015ej}
\bibinfo{author}{He, J.} \emph{et~al.}
\newblock \bibinfo{title}{{Ultracompact quantum splitter of degenerate photon
  pairs}}.
\newblock \emph{\bibinfo{journal}{Optica}} \textbf{\bibinfo{volume}{2}},
  \bibinfo{pages}{779--782} (\bibinfo{year}{2015}).

\bibitem{Davanco:2012et}
\bibinfo{author}{Davanco, M.} \emph{et~al.}
\newblock \bibinfo{title}{{Telecommunications-band heralded single photons from
  a silicon nanophotonic chip}}.
\newblock \emph{\bibinfo{journal}{Applied Physics Letters}}
  \textbf{\bibinfo{volume}{100}} (\bibinfo{year}{2012}).

\bibitem{Jiang:2012wt}
\bibinfo{author}{Jiang, W.~C.}, \bibinfo{author}{Lu, X.},
  \bibinfo{author}{Zhang, J.}, \bibinfo{author}{Painter, O.} \&
  \bibinfo{author}{Lin, Q.}
\newblock \bibinfo{title}{{A silicon-chip source of bright photon-pair comb}}.
\newblock \emph{\bibinfo{journal}{arXiv}} \bibinfo{pages}{1210.4455}
  (\bibinfo{year}{2012}).

\bibitem{ro-apl-107-041102}
\bibinfo{author}{Rogers, S.}, \bibinfo{author}{Lu, X.}, \bibinfo{author}{Jiang,
  W.~C.} \& \bibinfo{author}{Lin, Q.}
\newblock \bibinfo{title}{{Twin photon pairs in a high-Q silicon
  microresonator}}.
\newblock \emph{\bibinfo{journal}{Applied Physics Letters}}
  \textbf{\bibinfo{volume}{107}}, \bibinfo{pages}{041102+}
  (\bibinfo{year}{2015}).

\bibitem{Xiong:2011gt}
\bibinfo{author}{Xiong, C.} \emph{et~al.}
\newblock \bibinfo{title}{{Slow-light enhanced correlated photon pair
  generation in a silicon photonic crystal waveguide}}.
\newblock \emph{\bibinfo{journal}{Opt. Lett.}} \textbf{\bibinfo{volume}{36}},
  \bibinfo{pages}{3413--3415} (\bibinfo{year}{2011}).

\bibitem{Collins:2013eu}
\bibinfo{author}{Collins, M.~J.} \emph{et~al.}
\newblock \bibinfo{title}{{Integrated spatial multiplexing of heralded
  single-photon sources}}.
\newblock \emph{\bibinfo{journal}{Nature Communications}}
  \textbf{\bibinfo{volume}{4}}, \bibinfo{pages}{2582} (\bibinfo{year}{2013}).

\bibitem{Matsuda:2013cv}
\bibinfo{author}{Matsuda, N.} \emph{et~al.}
\newblock \bibinfo{title}{{Slow light enhanced correlated photon pair
  generation in photonic-crystal coupled-resonator optical waveguides.}}
\newblock \emph{\bibinfo{journal}{Op. Ex.}} \textbf{\bibinfo{volume}{21}},
  \bibinfo{pages}{8596--8604} (\bibinfo{year}{2013}).

\bibitem{Takesue:2014ic}
\bibinfo{author}{Takesue, H.}, \bibinfo{author}{Matsuda, N.},
  \bibinfo{author}{Kuramochi, E.} \& \bibinfo{author}{Notomi, M.}
\newblock \bibinfo{title}{{Entangled photons from on-chip slow light}}.
\newblock \emph{\bibinfo{journal}{Sci. Rep.}} \textbf{\bibinfo{volume}{4}},
  \bibinfo{pages}{3913} (\bibinfo{year}{2014}).

\bibitem{Azzini:1563418}
\bibinfo{author}{Azzini, S.} \emph{et~al.}
\newblock \bibinfo{title}{{Stimulated and spontaneous four-wave mixing in
  silicon-on-insulator coupled photonic wire nano-cavities}}.
\newblock \emph{\bibinfo{journal}{Applied Physics Letters}}
  \textbf{\bibinfo{volume}{103}}, \bibinfo{pages}{031117}
  (\bibinfo{year}{2013}).

\bibitem{Harris:2015ux}
\bibinfo{author}{Harris, N.~C.} \emph{et~al.}
\newblock \bibinfo{title}{{Bosonic transport simulations in a large-scale
  programmable nanophotonic processor}}.
\newblock \emph{\bibinfo{journal}{arXiv}} \bibinfo{pages}{1507.03406}
  (\bibinfo{year}{2015}).

\bibitem{Steinbrecher:2015ej}
\bibinfo{author}{Steinbrecher, G.}, \bibinfo{author}{Harris, N.~C.},
  \bibinfo{author}{Mower, J.}, \bibinfo{author}{Prabhu, M.} \&
  \bibinfo{author}{Englund, D.~R.}
\newblock \bibinfo{title}{{Programmable Nanophotonic Processor for Arbitrary
  High Fidelity Optical Transformations}}.
\newblock In \emph{\bibinfo{booktitle}{CLEO}}, \bibinfo{pages}{FW4A.2}
  (\bibinfo{publisher}{Optical Society of America},
  \bibinfo{address}{Washington, D.C.}, \bibinfo{year}{2015}).

\bibitem{Li:2015ue}
\bibinfo{author}{Li, Y.}, \bibinfo{author}{Humphreys, P.~C.},
  \bibinfo{author}{Mendoza, G.~J.} \& \bibinfo{author}{Benjamin, S.~C.}
\newblock \bibinfo{title}{{Resource Costs for Fault-Tolerant Linear Optical
  Quantum Computing}}.
\newblock \emph{\bibinfo{journal}{Phys. Rev. X}} \textbf{\bibinfo{volume}{5}},
  \bibinfo{pages}{041007} (\bibinfo{year}{2015}).

\bibitem{Carolan:2015fba}
\bibinfo{author}{Carolan, J.} \emph{et~al.}
\newblock \bibinfo{title}{{Universal linear optics}}.
\newblock \emph{\bibinfo{journal}{Science}} \textbf{\bibinfo{volume}{349}},
  \bibinfo{pages}{711--716} (\bibinfo{year}{2015}).

\bibitem{Aaronson:2011ho}
\bibinfo{author}{Aaronson, S.} \& \bibinfo{author}{Arkhipov, A.}
\newblock \emph{\bibinfo{title}{{The computational complexity of linear
  optics}}} (\bibinfo{publisher}{ACM}, \bibinfo{address}{New York, New York,
  USA}, \bibinfo{year}{2011}).

\bibitem{Tillmann:2013jva}
\bibinfo{author}{Tillmann, M.} \emph{et~al.}
\newblock \bibinfo{title}{{Experimental boson sampling}}.
\newblock \emph{\bibinfo{journal}{Nature Photonics}}
  \textbf{\bibinfo{volume}{7}}, \bibinfo{pages}{540--544}
  (\bibinfo{year}{2013}).

\bibitem{Bentivegna:2015ia}
\bibinfo{author}{Bentivegna, M.} \emph{et~al.}
\newblock \bibinfo{title}{{Experimental scattershot boson sampling}}.
\newblock \emph{\bibinfo{journal}{Science Advances}}
  \textbf{\bibinfo{volume}{1}}, \bibinfo{pages}{e1400255--e1400255}
  (\bibinfo{year}{2015}).

\bibitem{Hong:1987vi}
\bibinfo{author}{Hong, C.}, \bibinfo{author}{Ou, Z.} \&
  \bibinfo{author}{Mandel, L.}
\newblock \bibinfo{title}{{Measurement of subpicosecond time intervals between
  two photons by interference.}}
\newblock \emph{\bibinfo{journal}{Phys. Rev. Lett.}}
  \textbf{\bibinfo{volume}{59}}, \bibinfo{pages}{2044--2046}
  (\bibinfo{year}{1987}).

\bibitem{Bonneau:2012gx}
\bibinfo{author}{Bonneau, D.} \emph{et~al.}
\newblock \bibinfo{title}{{Quantum interference and manipulation of
  entanglement in silicon wire waveguide quantum circuits}}.
\newblock \emph{\bibinfo{journal}{New J. Phys.}} \textbf{\bibinfo{volume}{14}},
  \bibinfo{pages}{045003} (\bibinfo{year}{2012}).

\bibitem{Xu:2013jna}
\bibinfo{author}{Xu, X.} \emph{et~al.}
\newblock \bibinfo{title}{{Near-infrared Hong-Ou-Mandel interference on a
  silicon quantum photonic chip}}.
\newblock \emph{\bibinfo{journal}{Op. Ex.}} \textbf{\bibinfo{volume}{21}},
  \bibinfo{pages}{5014--5024} (\bibinfo{year}{2013}).

\bibitem{Silverstone:2015tl}
\bibinfo{author}{Silverstone, J.~W.} \emph{et~al.}
\newblock \bibinfo{title}{{Photonic qubit entanglement and processing in
  silicon waveguides.}}
\newblock \emph{\bibinfo{journal}{Advanced Photonics 2015 (2015)}}
  \bibinfo{pages}{IS4A.5} (\bibinfo{year}{2015}).

\bibitem{Najafi:2015ey}
\bibinfo{author}{Najafi, F.} \emph{et~al.}
\newblock \bibinfo{title}{{On-chip detection of non-classical light by scalable
  integration of single-photon detectors}}.
\newblock \emph{\bibinfo{journal}{Nature Communications}}
  \textbf{\bibinfo{volume}{6}}, \bibinfo{pages}{5873} (\bibinfo{year}{2015}).

\bibitem{Warburton:2013hk}
\bibinfo{author}{Warburton, R.~E.} \emph{et~al.}
\newblock \bibinfo{title}{{Ge-on-Si Single-Photon Avalanche Diode Detectors:
  Design, Modeling, Fabrication, and Characterization at Wavelengths 1310 and
  1550 nm}}.
\newblock \emph{\bibinfo{journal}{Electron Devices, IEEE Transactions on}}
  \textbf{\bibinfo{volume}{60}}, \bibinfo{pages}{3807--3813}
  (\bibinfo{year}{2013}).

\bibitem{Kimble:1977iw}
\bibinfo{author}{Kimble, H.~J.}, \bibinfo{author}{Dagenais, M.} \&
  \bibinfo{author}{Mandel, L.}
\newblock \bibinfo{title}{{Photon Antibunching in Resonance Fluorescence}}.
\newblock \emph{\bibinfo{journal}{Phys. Rev. Lett.}}
  \textbf{\bibinfo{volume}{39}}, \bibinfo{pages}{691--695}
  (\bibinfo{year}{1977}).

\bibitem{Aspect:1981ga}
\bibinfo{author}{Aspect, A.}, \bibinfo{author}{Aspect, A.},
  \bibinfo{author}{Grangier, P.}, \bibinfo{author}{Grangier, P.} \&
  \bibinfo{author}{Roger, G.}
\newblock \bibinfo{title}{{Experimental Tests of Realistic Local Theories via
  Bell's Theorem}}.
\newblock \emph{\bibinfo{journal}{Phys. Rev. Lett.}}
  \textbf{\bibinfo{volume}{47}}, \bibinfo{pages}{460--463}
  (\bibinfo{year}{1981}).

\bibitem{Goltsman:2001ea}
\bibinfo{author}{Okunev, O.} \emph{et~al.}
\newblock \bibinfo{title}{{Picosecond superconducting single-photon optical
  detector}}.
\newblock \emph{\bibinfo{journal}{Applied Physics Letters}}
  \textbf{\bibinfo{volume}{79}}, \bibinfo{pages}{705--707}
  (\bibinfo{year}{2001}).

\bibitem{Dauler:2014bq}
\bibinfo{author}{Dauler, E.~A.} \emph{et~al.}
\newblock \bibinfo{title}{{Review of superconducting nanowire single-photon
  detector system design options and demonstrated performance}}.
\newblock \emph{\bibinfo{journal}{Optical Engineering}}
  \textbf{\bibinfo{volume}{53}}, \bibinfo{pages}{081907--081907}
  (\bibinfo{year}{2014}).

\bibitem{Heath:2015da}
\bibinfo{author}{Heath, R.~M.} \emph{et~al.}
\newblock \bibinfo{title}{{Nanoantenna Enhancement for Telecom-Wavelength
  Superconducting Single Photon Detectors}}.
\newblock \emph{\bibinfo{journal}{Nano Lett.}} \textbf{\bibinfo{volume}{15}},
  \bibinfo{pages}{819--822} (\bibinfo{year}{2015}).

\bibitem{Rath:2015um}
\bibinfo{author}{Rath, P.} \emph{et~al.}
\newblock \bibinfo{title}{{Superconducting single photon detectors integrated
  with diamond nanophotonic circuits}}.
\newblock \emph{\bibinfo{journal}{arXiv}} \bibinfo{pages}{1505.04251}
  (\bibinfo{year}{2015}).

\bibitem{Marsili:2012th}
\bibinfo{author}{Marsili, F.} \emph{et~al.}
\newblock \bibinfo{title}{{Detecting single infrared photons with 93\% system
  efficiency}}.
\newblock \emph{\bibinfo{journal}{Nature Photonics}}
  \textbf{\bibinfo{volume}{7}}, \bibinfo{pages}{210--214}
  (\bibinfo{year}{2013}).

\bibitem{Verma:2014dw}
\bibinfo{author}{Verma, V.~B.} \emph{et~al.}
\newblock \bibinfo{title}{{High-efficiency WSi superconducting nanowire
  single-photon detectors operating at 2.5 K}}.
\newblock \emph{\bibinfo{journal}{Applied Physics Letters}}
  \textbf{\bibinfo{volume}{105}}, \bibinfo{pages}{122601}
  (\bibinfo{year}{2014}).

\bibitem{Beyer:2015ga}
\bibinfo{author}{Beyer, A.} \emph{et~al.}
\newblock \bibinfo{title}{{Waveguide-Coupled Superconducting Nanowire
  Single-Photon Detectors}}.
\newblock \emph{\bibinfo{journal}{CLEO: 2015}} \bibinfo{pages}{STh1I.2}
  (\bibinfo{year}{2015}).

\bibitem{Verma:2015vg}
\bibinfo{author}{Verma, V.~B.} \emph{et~al.}
\newblock \bibinfo{title}{{High-efficiency superconducting nanowire
  single-photon detectors fabricated from MoSi thin-films}}.
\newblock \emph{\bibinfo{journal}{arXiv}} \bibinfo{pages}{1504.02793}
  (\bibinfo{year}{2015}).

\bibitem{Korneeva:2014es}
\bibinfo{author}{Korneeva, Y.~P.} \emph{et~al.}
\newblock \bibinfo{title}{{Superconducting single-photon detector made of MoSi
  film}}.
\newblock \emph{\bibinfo{journal}{Supercond. Sci. Technol.}}
  \textbf{\bibinfo{volume}{27}}, \bibinfo{pages}{095012}
  (\bibinfo{year}{2014}).

\bibitem{Verma:2014ez}
\bibinfo{author}{Verma, V.~B.} \emph{et~al.}
\newblock \bibinfo{title}{{Superconducting nanowire single photon detectors
  fabricated from an amorphous Mo$_{0.75}$Ge$_{0.25}$ thin film}}.
\newblock \emph{\bibinfo{journal}{Applied Physics Letters}}
  \textbf{\bibinfo{volume}{105}}, \bibinfo{pages}{022602}
  (\bibinfo{year}{2014}).

\bibitem{Pernice:2012bc}
\bibinfo{author}{Pernice, W. H.~P.} \emph{et~al.}
\newblock \bibinfo{title}{{High-speed and high-efficiency travelling wave
  single-photon detectors embedded in nanophotonic circuits}}.
\newblock \emph{\bibinfo{journal}{Nature Communications}}
  \textbf{\bibinfo{volume}{3}}, \bibinfo{pages}{1325} (\bibinfo{year}{2012}).

\bibitem{Sprengers:2011era}
\bibinfo{author}{Sprengers, J.~P.} \emph{et~al.}
\newblock \bibinfo{title}{{Waveguide superconducting single-photon detectors
  for integrated quantum photonic circuits}}.
\newblock \emph{\bibinfo{journal}{Applied Physics Letters}}
  \textbf{\bibinfo{volume}{99}}, \bibinfo{pages}{181110--181110}
  (\bibinfo{year}{2011}).

\bibitem{Reithmaier:2013jk}
\bibinfo{author}{Reithmaier, G.} \emph{et~al.}
\newblock \bibinfo{title}{{On-chip time resolved detection of quantum dot
  emission using integrated superconducting single photon detectors}}.
\newblock \emph{\bibinfo{journal}{Sci. Rep.}} \textbf{\bibinfo{volume}{3}},
  \bibinfo{pages}{1901} (\bibinfo{year}{2013}).

\bibitem{KLM2001}
\bibinfo{author}{Knill, E.}, \bibinfo{author}{Laflamme, R.} \&
  \bibinfo{author}{Milburn, G.~J.}
\newblock \bibinfo{title}{{A scheme for efficient quantum computation with
  linear optics}}.
\newblock \emph{\bibinfo{journal}{Nature}} \textbf{\bibinfo{volume}{409}},
  \bibinfo{pages}{46--52} (\bibinfo{year}{2001}).

\bibitem{ha-oe-22-10487}
\bibinfo{author}{Harris, N.~C.} \emph{et~al.}
\newblock \bibinfo{title}{{Efficient, compact and low loss thermo-optic phase
  shifter in silicon}}.
\newblock \emph{\bibinfo{journal}{Optics Express}}
  \textbf{\bibinfo{volume}{22}}, \bibinfo{pages}{10487+}
  (\bibinfo{year}{2014}).

\bibitem{wa-oe-38-733}
\bibinfo{author}{Watts, M.~R.} \emph{et~al.}
\newblock \bibinfo{title}{{Adiabatic thermo-optic Machâ€“Zehnder switch}}.
\newblock \emph{\bibinfo{journal}{Optics Letters}}
  \textbf{\bibinfo{volume}{38}}, \bibinfo{pages}{733+} (\bibinfo{year}{2013}).

\bibitem{se-optica-3-64}
\bibinfo{author}{Seok, T.~J.}, \bibinfo{author}{Quack, N.},
  \bibinfo{author}{Han, S.}, \bibinfo{author}{Muller, R.~S.} \&
  \bibinfo{author}{Wu, M.~C.}
\newblock \bibinfo{title}{{Large-scale broadband digital silicon photonic
  switches with vertical adiabatic couplers}}.
\newblock \emph{\bibinfo{journal}{Optica}} \textbf{\bibinfo{volume}{3}},
  \bibinfo{pages}{64+} (\bibinfo{year}{2016}).

\bibitem{lithium_niobate_reviewIEEE}
\bibinfo{author}{Wooten, E.~L.} \emph{et~al.}
\newblock \bibinfo{title}{{A review of lithium niobate modulators for
  fiber-optic communications systems}}.
\newblock \emph{\bibinfo{journal}{Selected Topics in Quantum Electronics, IEEE
  Journal of}} \textbf{\bibinfo{volume}{6}}, \bibinfo{pages}{69--82}
  (\bibinfo{year}{2000}).

\bibitem{Predevelclusterstate2007}
\bibinfo{author}{Prevedel, R.} \emph{et~al.}
\newblock \bibinfo{title}{{High-speed linear optics quantum computing using
  active feed-forward}}.
\newblock \emph{\bibinfo{journal}{Nature}} \textbf{\bibinfo{volume}{445}},
  \bibinfo{pages}{65--69} (\bibinfo{year}{2007}).

\bibitem{bo-prl-ln-2012}
\bibinfo{author}{Bonneau, D.} \emph{et~al.}
\newblock \bibinfo{title}{{Fast Path and Polarization Manipulation of Telecom
  Wavelength Single Photons in Lithium Niobate Waveguide Devices}}.
\newblock \emph{\bibinfo{journal}{Physical Review Letters}}
  \textbf{\bibinfo{volume}{108}}, \bibinfo{pages}{053601+}
  (\bibinfo{year}{2012}).

\bibitem{th-np-4-518}
\bibinfo{author}{Reed, G.~T.}, \bibinfo{author}{Mashanovich, G.},
  \bibinfo{author}{Gardes, F.~Y.} \& \bibinfo{author}{Thomson, D.~J.}
\newblock \bibinfo{title}{{Silicon optical modulators}}.
\newblock \emph{\bibinfo{journal}{Nat Photon}} \textbf{\bibinfo{volume}{4}},
  \bibinfo{pages}{518--526} (\bibinfo{year}{2010}).

\bibitem{bu-prb-11-5848}
\bibinfo{author}{Bude, J.}, \bibinfo{author}{Sano, N.} \&
  \bibinfo{author}{Yoshii, A.}
\newblock \bibinfo{title}{{Hot-carrier luminescence in Si}}.
\newblock \emph{\bibinfo{journal}{Physical Review B}}
  \textbf{\bibinfo{volume}{45}}, \bibinfo{pages}{5848--5856}
  (\bibinfo{year}{1992}).

\bibitem{SAIEE-103-18}
\bibinfo{author}{Bogalecki, A.~W.}, \bibinfo{author}{du~Plessis, M.},
  \bibinfo{author}{Venter, P.~J.} \& \bibinfo{author}{van Rensburg, C.~J.}
\newblock \bibinfo{title}{Spectral measurement and analysis of silicon cmos
  light sources}.
\newblock vol. \bibinfo{volume}{103}, \bibinfo{pages}{18--23}
  (\bibinfo{year}{2012}).

\bibitem{Al-nat-431-1081}
\bibinfo{author}{Almeida, V.~R.}, \bibinfo{author}{Barrios, C.~A.},
  \bibinfo{author}{Panepucci, R.~R.} \& \bibinfo{author}{Lipson, M.}
\newblock \bibinfo{title}{{All-optical control of light on a silicon chip}}.
\newblock \emph{\bibinfo{journal}{Nature}} \textbf{\bibinfo{volume}{431}},
  \bibinfo{pages}{1081--1084} (\bibinfo{year}{2004}).

\bibitem{so64374}
\bibinfo{author}{{Dekker}, R.}, \bibinfo{author}{{Niehusmann}, J.},
  \bibinfo{author}{{F\"{o}rst}, M.} \& \bibinfo{author}{{Driessen}, A.}
\newblock \bibinfo{title}{{Ultrafast all-optical switching by cross phase
  modulation induced wavelength conversion in silicon-on-insulator waveguides
  and ring resonators}}.
\newblock In \emph{\bibinfo{booktitle}{European Conference on Integrated
  Optics, ECIO 2007}} (\bibinfo{year}{2007}).

\bibitem{Hall:2011ki}
\bibinfo{author}{Hall, M.~A.}, \bibinfo{author}{Altepeter, J.~B.} \&
  \bibinfo{author}{Kumar, P.}
\newblock \bibinfo{title}{{All-optical switching of photonic entanglement}}.
\newblock \emph{\bibinfo{journal}{New J. Phys.}} \textbf{\bibinfo{volume}{13}},
  \bibinfo{pages}{105004} (\bibinfo{year}{2011}).

\bibitem{Nature2006SiliconKhi2}
\bibinfo{author}{Jacobsen, R.~S.} \emph{et~al.}
\newblock \bibinfo{title}{{Strained silicon as a new electro-optic material}}.
\newblock \emph{\bibinfo{journal}{Nature}} \textbf{\bibinfo{volume}{441}},
  \bibinfo{pages}{199--202} (\bibinfo{year}{2006}).

\bibitem{Timurdogan:2016wh}
\bibinfo{author}{Timurdogan, E.}, \bibinfo{author}{Poulton, C.~V.} \&
  \bibinfo{author}{Watts, M.~R.}
\newblock \bibinfo{title}{{Electric Field-Induced Second Order Nonlinear
  Optical Effects in Silicon Waveguides}}.
\newblock \emph{\bibinfo{journal}{arXiv}} \bibinfo{pages}{1603.04515}
  (\bibinfo{year}{2016}).

\bibitem{Xiong-AlN:2012}
\bibinfo{author}{Xiong, C.}, \bibinfo{author}{Pernice, W. H.~P.} \&
  \bibinfo{author}{Tang, H.~X.}
\newblock \bibinfo{title}{{Low-Loss, Silicon Integrated, Aluminum Nitride
  Photonic Circuits and Their Use for Electro-Optic Signal Processing}}.
\newblock \emph{\bibinfo{journal}{Nano Lett.}} \textbf{\bibinfo{volume}{12}},
  \bibinfo{pages}{3562--3568} (\bibinfo{year}{2012}).

\bibitem{Al-OptExp-20110606}
\bibinfo{author}{Alloatti, L.} \emph{et~al.}
\newblock \bibinfo{title}{{42.7 Gbit/s electro-optic modulator in silicon
  technology}}.
\newblock \emph{\bibinfo{journal}{Opt. Express}} \textbf{\bibinfo{volume}{19}},
  \bibinfo{pages}{11841--11851} (\bibinfo{year}{2011}).

\bibitem{Ab-natcomm-4-1671}
\bibinfo{author}{Abel, S.} \emph{et~al.}
\newblock \bibinfo{title}{{A strong electro-optically active lead-free
  ferroelectric integrated on silicon}}.
\newblock \emph{\bibinfo{journal}{Nature Communications}}
  \textbf{\bibinfo{volume}{4}}, \bibinfo{pages}{1671+} (\bibinfo{year}{2013}).

\bibitem{Xi-nl-14-1419}
\bibinfo{author}{Xiong, C.} \emph{et~al.}
\newblock \bibinfo{title}{{Active Silicon Integrated Nanophotonics:
  Ferroelectric BaTiO3 Devices}}.
\newblock \emph{\bibinfo{journal}{Nano Lett.}} \textbf{\bibinfo{volume}{14}},
  \bibinfo{pages}{1419--1425} (\bibinfo{year}{2014}).

\bibitem{Abel:2013fj}
\bibinfo{author}{Abel, S.} \emph{et~al.}
\newblock \bibinfo{title}{{Electro-Optical Active Barium Titanate Thin Films in
  Silicon Photonics Devices}}.
\newblock \emph{\bibinfo{journal}{IPR}}  (\bibinfo{year}{2013}).

\bibitem{Ong:2013uh}
\bibinfo{author}{Ong, J.~R.}, \bibinfo{author}{Kumar, R.} \&
  \bibinfo{author}{Mookherjea, S.}
\newblock \bibinfo{title}{{Ultra-High-Contrast and Tunable-Bandwidth Filter
  Using Cascaded High-Order Silicon Microring Filters}}.
\newblock \emph{\bibinfo{journal}{Photonics Technology Letters, IEEE}}
  \textbf{\bibinfo{volume}{25}}, \bibinfo{pages}{1543--1546}
  (\bibinfo{year}{2013}).

\bibitem{mat-filter2chips-2016}
\bibinfo{author}{Piekarek, M.} \emph{et~al.}
\newblock \bibinfo{title}{Passive high-extinction integrated photonic filters
  for silicon quantum photonics}.
\newblock \emph{\bibinfo{journal}{in preparation}}  (\bibinfo{year}{2016}).

\bibitem{Matsuda:2014cy}
\bibinfo{author}{Matsuda, N.} \emph{et~al.}
\newblock \bibinfo{title}{{On-chip generation and demultiplexing of quantum
  correlated photons using a silicon-silica monolithic photonic integration
  platform.}}
\newblock \emph{\bibinfo{journal}{Op. Ex.}} \textbf{\bibinfo{volume}{22}},
  \bibinfo{pages}{22831--22840} (\bibinfo{year}{2014}).

\bibitem{Almeida:2003tp}
\bibinfo{author}{Almeida, V.~R.}, \bibinfo{author}{Panepucci, R.~R.} \&
  \bibinfo{author}{Lipson, M.}
\newblock \bibinfo{title}{{Nanotaper for compact mode conversion}}.
\newblock \emph{\bibinfo{journal}{Optics Letters}}
  \textbf{\bibinfo{volume}{28}}, \bibinfo{pages}{1302--1304}
  (\bibinfo{year}{2003}).

\bibitem{ro-oe-10-1364}
\bibinfo{author}{Roelkens, G.}, \bibinfo{author}{Thourhout, D.~V.} \&
  \bibinfo{author}{Baets, R.}
\newblock \bibinfo{title}{{High efficiency grating coupler between
  silicon-on-insulator waveguides and perfectly vertical optical fibers}}.
\newblock \emph{\bibinfo{journal}{Optics Letters}}
  \textbf{\bibinfo{volume}{32}}, \bibinfo{pages}{1495+} (\bibinfo{year}{2007}).

\bibitem{Vivien:06}
\bibinfo{author}{Vivien, L.} \emph{et~al.}
\newblock \bibinfo{title}{{Light Injection in SOI Microwaveguides Using
  High-Efficiency Grating Couplers}}.
\newblock \emph{\bibinfo{journal}{J. Lightwave Technol.}}
  \textbf{\bibinfo{volume}{24}}, \bibinfo{pages}{3810--3815}
  (\bibinfo{year}{2006}).

\bibitem{Cardenas:2014tp}
\bibinfo{author}{Cardenas, J.} \emph{et~al.}
\newblock \bibinfo{title}{{High Coupling Efficiency Etched Facet Tapers in
  Silicon Waveguides}}.
\newblock \emph{\bibinfo{journal}{Photonics Technology Letters, IEEE}}
  \textbf{\bibinfo{volume}{26}}, \bibinfo{pages}{2380--2382}
  (\bibinfo{year}{2014}).

\bibitem{Notaros:2016vw}
\bibinfo{author}{Notaros, J.} \emph{et~al.}
\newblock \bibinfo{title}{{Ultra-Efficient CMOS Fiber-to-Chip Grating
  Couplers}}.
\newblock In \emph{\bibinfo{booktitle}{OFC: 2016}}, \bibinfo{pages}{M2I.5}
  (\bibinfo{publisher}{Optical Society of America},
  \bibinfo{address}{Washington, D.C.}, \bibinfo{year}{2016}).

\bibitem{Taillaert:2003gc}
\bibinfo{author}{Taillaert, D.} \emph{et~al.}
\newblock \bibinfo{title}{{A compact two-dimensional grating coupler used as a
  polarization splitter}}.
\newblock \emph{\bibinfo{journal}{Photonics Technology Letters, IEEE}}
  \textbf{\bibinfo{volume}{15}}, \bibinfo{pages}{1249--1251}
  (\bibinfo{year}{2003}).

\bibitem{Wang:2016eo}
\bibinfo{author}{Wang, J.} \emph{et~al.}
\newblock \bibinfo{title}{{Chip-to-chip quantum photonic interconnect by
  path-polarization interconversion}}.
\newblock \emph{\bibinfo{journal}{Optica}} \textbf{\bibinfo{volume}{3}},
  \bibinfo{pages}{407--413} (\bibinfo{year}{2016}).

\bibitem{Harada2008Generation}
\bibinfo{author}{Harada, K.-i.} \emph{et~al.}
\newblock \bibinfo{title}{{Generation of high-purity entangled photon pairs
  using silicon wire waveguide}}.
\newblock \emph{\bibinfo{journal}{Optics Express}}
  \textbf{\bibinfo{volume}{16}}, \bibinfo{pages}{20368--20373}
  (\bibinfo{year}{2008}).

\bibitem{ra-prl-22-5188}
\bibinfo{author}{Raussendorf, R.} \& \bibinfo{author}{Briegel, H.~J.}
\newblock \bibinfo{title}{{A One-Way Quantum Computer}}.
\newblock \emph{\bibinfo{journal}{Physical Review Letters}}
  \textbf{\bibinfo{volume}{86}}, \bibinfo{pages}{5188--5191}
  (\bibinfo{year}{2001}).

\bibitem{Ly-np-3-706}
\bibinfo{author}{Lvovsky, A.~I.}, \bibinfo{author}{Sanders, B.~C.} \&
  \bibinfo{author}{Tittel, W.}
\newblock \bibinfo{title}{{Optical quantum memory}}.
\newblock \emph{\bibinfo{journal}{Nature Photonics}}
  \textbf{\bibinfo{volume}{3}}, \bibinfo{pages}{706--714}
  (\bibinfo{year}{2009}).

\bibitem{be-delay-icton-2013}
\bibinfo{author}{Beggs, D.~M.} \emph{et~al.}
\newblock \bibinfo{title}{{Optical delay in silicon photonic crystals using
  ultrafast indirect photonic transitions}}.
\newblock In \emph{\bibinfo{booktitle}{Transparent Optical Networks (ICTON),
  2013 15th International Conference on}}, \bibinfo{pages}{1--4}
  (\bibinfo{publisher}{IEEE}, \bibinfo{year}{2013}).

\bibitem{kh-oe-19-11780}
\bibinfo{author}{Khan, S.}, \bibinfo{author}{Baghban, M.~A.} \&
  \bibinfo{author}{Fathpour, S.}
\newblock \bibinfo{title}{{Electronically tunable silicon photonic delay
  lines}}.
\newblock \emph{\bibinfo{journal}{Optics Express}}
  \textbf{\bibinfo{volume}{19}}, \bibinfo{pages}{11780+}
  (\bibinfo{year}{2011}).

\bibitem{melloni-delayline-2010}
\bibinfo{author}{Melloni, A.} \emph{et~al.}
\newblock \bibinfo{title}{{Tunable Delay Lines in Silicon Photonics: Coupled
  Resonators and Photonic Crystals, a Comparison}}.
\newblock \emph{\bibinfo{journal}{Photonics Journal, IEEE}}
  \textbf{\bibinfo{volume}{2}}, \bibinfo{pages}{181--194}
  (\bibinfo{year}{2010}).

\bibitem{po-josab-21-1665}
\bibinfo{author}{Poon, J. K.~S.}, \bibinfo{author}{Scheuer, J.},
  \bibinfo{author}{Xu, Y.} \& \bibinfo{author}{Yariv, A.}
\newblock \bibinfo{title}{{Designing coupled-resonator optical waveguide delay
  lines}}.
\newblock \emph{\bibinfo{journal}{Journal of the Optical Society of America B}}
  \textbf{\bibinfo{volume}{21}}, \bibinfo{pages}{1665--1673}
  (\bibinfo{year}{2004}).

\bibitem{Takesue:2013db}
\bibinfo{author}{Takesue, H.}, \bibinfo{author}{Matsuda, N.},
  \bibinfo{author}{Kuramochi, E.}, \bibinfo{author}{Munro, W.~J.} \&
  \bibinfo{author}{Notomi, M.}
\newblock \bibinfo{title}{{An on-chip coupled resonator optical waveguide
  single-photon buffer}}.
\newblock \emph{\bibinfo{journal}{Nature Communications}}
  \textbf{\bibinfo{volume}{4}} (\bibinfo{year}{2013}).

\bibitem{xi-oe-22-22707}
\bibinfo{author}{Xie, J.}, \bibinfo{author}{Zhou, L.}, \bibinfo{author}{Li,
  Z.}, \bibinfo{author}{Wang, J.} \& \bibinfo{author}{Chen, J.}
\newblock \bibinfo{title}{{Seven-bit reconfigurable optical true time delay
  line based on silicon integration}}.
\newblock \emph{\bibinfo{journal}{Optics Express}}
  \textbf{\bibinfo{volume}{22}}, \bibinfo{pages}{22707+}
  (\bibinfo{year}{2014}).

\bibitem{az-ncomm-6-6787}
\bibinfo{author}{Azuma, K.}, \bibinfo{author}{Tamaki, K.} \&
  \bibinfo{author}{Lo, H.-K.}
\newblock \bibinfo{title}{{All-photonic quantum repeaters}}.
\newblock \emph{\bibinfo{journal}{Nature Communications}}
  \textbf{\bibinfo{volume}{6}}, \bibinfo{pages}{6787+} (\bibinfo{year}{2015}).

\bibitem{GimenoSegovia:2014uf}
\bibinfo{author}{Gimeno-Segovia, M.}, \bibinfo{author}{Shadbolt, P.} \&
  \bibinfo{author}{Browne, D.~E.}
\newblock \bibinfo{title}{{From Three-Photon Greenberger-Horne-Zeilinger States
  to Ballistic Universal Quantum Computation}}.
\newblock \emph{\bibinfo{journal}{Phys. Rev. Lett.}}
  \textbf{\bibinfo{volume}{115}}, \bibinfo{pages}{020502}
  (\bibinfo{year}{2015}).

\bibitem{lee-natcom-3-867}
\bibinfo{author}{Lee, H.}, \bibinfo{author}{Chen, T.}, \bibinfo{author}{Li,
  J.}, \bibinfo{author}{Painter, O.} \& \bibinfo{author}{Vahala, K.~J.}
\newblock \bibinfo{title}{{Ultra-low-loss optical delay line on a silicon
  chip}}.
\newblock \emph{\bibinfo{journal}{Nature Communications}}
  \textbf{\bibinfo{volume}{3}}, \bibinfo{pages}{867+} (\bibinfo{year}{2012}).

\bibitem{ba-oe-21-544}
\bibinfo{author}{Bauters, J.~F.} \emph{et~al.}
\newblock \bibinfo{title}{{Silicon on ultra-low-loss waveguide photonic
  integration platform}}.
\newblock \emph{\bibinfo{journal}{Opt. Express}} \textbf{\bibinfo{volume}{21}},
  \bibinfo{pages}{544--555} (\bibinfo{year}{2013}).

\bibitem{Vlasov:2004bo}
\bibinfo{author}{Vlasov, Y.~A.} \& \bibinfo{author}{McNab, S.~J.}
\newblock \bibinfo{title}{{Losses in single-mode silicon-on-insulator strip
  waveguides and bends}}.
\newblock \emph{\bibinfo{journal}{Op. Ex.}} \textbf{\bibinfo{volume}{12}},
  \bibinfo{pages}{1622--1631} (\bibinfo{year}{2004}).

\bibitem{Ou:2003ci}
\bibinfo{author}{Ou, H.}
\newblock \bibinfo{title}{{Different index contrast silica-on-silicon
  waveguides by PECVD}}.
\newblock \emph{\bibinfo{journal}{Electron Lett}}
  \textbf{\bibinfo{volume}{39}}, \bibinfo{pages}{212--213}
  (\bibinfo{year}{2003}).

\bibitem{Crespi:2011cy}
\bibinfo{author}{Crespi, A.} \emph{et~al.}
\newblock \bibinfo{title}{{Integrated photonic quantum gates for polarization
  qubits}}.
\newblock \emph{\bibinfo{journal}{Nature Communications}}
  \textbf{\bibinfo{volume}{2}}, \bibinfo{pages}{566} (\bibinfo{year}{2011}).

\bibitem{Moss:2013kv}
\bibinfo{author}{Moss, D.~J.}, \bibinfo{author}{Morandotti, R.},
  \bibinfo{author}{Gaeta, A.~L.} \& \bibinfo{author}{Lipson, M.}
\newblock \bibinfo{title}{{New CMOS-compatible platforms based on silicon
  nitride and Hydex for nonlinear optics}}.
\newblock \emph{\bibinfo{journal}{Nature Photonics}}
  \textbf{\bibinfo{volume}{7}}, \bibinfo{pages}{597--607}
  (\bibinfo{year}{2013}).

\bibitem{Van:2001je}
\bibinfo{author}{Van, V.}, \bibinfo{author}{Absil, P.~P.},
  \bibinfo{author}{Hryniewicz, J.~V.} \& \bibinfo{author}{Ho, P.~T.}
\newblock \bibinfo{title}{{Propagation loss in single-mode GaAs-AlGaAs
  microring resonators: measurement and model}}.
\newblock \emph{\bibinfo{journal}{J. Lightwave Technol.}}
  \textbf{\bibinfo{volume}{19}}, \bibinfo{pages}{1734--1739}
  (\bibinfo{year}{2001}).

\bibitem{Marpaung:2013gq}
\bibinfo{author}{Marpaung, D.} \emph{et~al.}
\newblock \bibinfo{title}{{Integrated microwave photonics}}.
\newblock \emph{\bibinfo{journal}{Laser {\&} Photonics Reviews}}
  \textbf{\bibinfo{volume}{7}}, \bibinfo{pages}{506--538}
  (\bibinfo{year}{2013}).

\bibitem{Ramelow:2015dx}
\bibinfo{author}{Ramelow, S.} \emph{et~al.}
\newblock \bibinfo{title}{{Monolithic Source of Tunable Narrowband Photons for
  Future Quantum Networks}}.
\newblock In \emph{\bibinfo{booktitle}{CLEO}}, \bibinfo{pages}{FM2A.7}
  (\bibinfo{publisher}{Optical Society of America},
  \bibinfo{address}{Washington, D.C.}, \bibinfo{year}{2015}).

\bibitem{Hiemstra:2015ue}
\bibinfo{author}{Hiemstra, T.} \emph{et~al.}
\newblock \bibinfo{title}{{Generating Telecom-Band Pure Heralded Single Photons
  On A Silica Chip}}.
\newblock \emph{\bibinfo{journal}{CLEO Europe: 2015}}
  \bibinfo{pages}{JSV--P--6} (\bibinfo{year}{2015}).

\bibitem{Spring:2016ti}
\bibinfo{author}{Spring, J.~B.} \emph{et~al.}
\newblock \bibinfo{title}{{A chip-based array of near-identical, pure, heralded
  single photon sources}}.
\newblock \emph{\bibinfo{journal}{arXiv}} \bibinfo{pages}{1603.06984}
  (\bibinfo{year}{2016}).

\bibitem{Spring:2013eg}
\bibinfo{author}{Spring, J.~B.} \emph{et~al.}
\newblock \bibinfo{title}{{On-chip low loss heralded source of pure single
  photons}}.
\newblock \emph{\bibinfo{journal}{Op. Ex.}} \textbf{\bibinfo{volume}{21}},
  \bibinfo{pages}{13522--13532} (\bibinfo{year}{2013}).

\bibitem{Autebert:2015cn}
\bibinfo{author}{Autebert, C.} \emph{et~al.}
\newblock \bibinfo{title}{{Photon pair sources in AlGaAs: from electrical
  injection to quantum state engineering}}.
\newblock \emph{\bibinfo{journal}{Journal of Modern Optics}}
  \textbf{\bibinfo{volume}{62}}, \bibinfo{pages}{1739--1745}
  (\bibinfo{year}{2015}).

\bibitem{Orieux:2013jv}
\bibinfo{author}{Orieux, A.} \emph{et~al.}
\newblock \bibinfo{title}{{Direct Bell States Generation on a III-V
  Semiconductor Chip at Room Temperature}}.
\newblock \emph{\bibinfo{journal}{Phys. Rev. Lett.}}
  \textbf{\bibinfo{volume}{110}}, \bibinfo{pages}{160502}
  (\bibinfo{year}{2013}).

\bibitem{Horn:2012}
\bibinfo{author}{Horn, R.} \emph{et~al.}
\newblock \bibinfo{title}{{Monolithic Source of Photon Pairs}}.
\newblock \emph{\bibinfo{journal}{Physical Review Letters}}
  \textbf{\bibinfo{volume}{108}}, \bibinfo{pages}{153605}
  (\bibinfo{year}{2012}).

\bibitem{Tanzilli:2011hk}
\bibinfo{author}{Tanzilli, S.} \emph{et~al.}
\newblock \bibinfo{title}{{On the genesis and evolution of Integrated Quantum
  Optics}}.
\newblock \emph{\bibinfo{journal}{Laser {\&} Photonics Reviews}}
  \textbf{\bibinfo{volume}{6}}, \bibinfo{pages}{115--143}
  (\bibinfo{year}{2011}).

\bibitem{Claudon:2010cb}
\bibinfo{author}{Claudon, J.} \emph{et~al.}
\newblock \bibinfo{title}{{A highly efficient single-photon source based on a
  quantum dot in a photonic nanowire}}.
\newblock \emph{\bibinfo{journal}{Nature Photonics}}
  \textbf{\bibinfo{volume}{4}}, \bibinfo{pages}{174--177}
  (\bibinfo{year}{2010}).

\bibitem{Sun:2013db}
\bibinfo{author}{Sun, X.}, \bibinfo{author}{Zhang, X.},
  \bibinfo{author}{Schuck, C.} \& \bibinfo{author}{Tang, H.~X.}
\newblock \bibinfo{title}{{Nonlinear optical effects of ultrahigh-Q silicon
  photonic nanocavities immersed in superfluid helium}}.
\newblock \emph{\bibinfo{journal}{Sci. Rep.}} \textbf{\bibinfo{volume}{3}}
  (\bibinfo{year}{2013}).

\bibitem{Jacobsen:2006up}
\bibinfo{author}{Jacobsen, R.~S.} \emph{et~al.}
\newblock \bibinfo{title}{{Strained silicon as a new electro-optic material}}.
\newblock \emph{\bibinfo{journal}{Nature}} \textbf{\bibinfo{volume}{441}},
  \bibinfo{pages}{199--202} (\bibinfo{year}{2006}).

\bibitem{Komma:2012dj}
\bibinfo{author}{Komma, J.}, \bibinfo{author}{Schwarz, C.},
  \bibinfo{author}{Hofmann, G.}, \bibinfo{author}{Heinert, D.} \&
  \bibinfo{author}{Nawrodt, R.}
\newblock \bibinfo{title}{{Thermo-optic coefficient of silicon at 1550 nm and
  cryogenic temperatures}}.
\newblock \emph{\bibinfo{journal}{Applied Physics Letters}}
  \textbf{\bibinfo{volume}{101}}, \bibinfo{pages}{041905}
  (\bibinfo{year}{2012}).

\bibitem{Ekanayake:2010fj}
\bibinfo{author}{Ekanayake, S.~R.}, \bibinfo{author}{Lehmann, T.},
  \bibinfo{author}{Dzurak, A.~S.}, \bibinfo{author}{Clark, R.~G.} \&
  \bibinfo{author}{Brawley, A.}
\newblock \bibinfo{title}{{Characterization of SOS-CMOS FETs at Low
  Temperatures for the Design of Integrated Circuits for Quantum Bit Control
  and Readout}}.
\newblock \emph{\bibinfo{journal}{Electron Devices, IEEE Transactions on}}
  \textbf{\bibinfo{volume}{57}}, \bibinfo{pages}{539--547}
  (\bibinfo{year}{2010}).

\bibitem{NAGATA:2011hj}
\bibinfo{author}{Nagata, H.} \emph{et~al.}
\newblock \bibinfo{title}{{Development of Cryogenic Readout Electronics for
  Far-Infrared Astronomical Focal Plane Array}}.
\newblock \emph{\bibinfo{journal}{IEICE Trans. Commun.}}
  \textbf{\bibinfo{volume}{E94-B}}, \bibinfo{pages}{2952--2960}
  (\bibinfo{year}{2011}).

\bibitem{Anonymous:1974tr}
\bibinfo{author}{Lengeler, B.}
\newblock \bibinfo{title}{{Semiconductor devices suitable for use in cryogenic
  environments}}.
\newblock \emph{\bibinfo{journal}{Cryogenics}} \bibinfo{pages}{439--447}
  (\bibinfo{year}{1974}).

\bibitem{Likharev:1991dg}
\bibinfo{author}{Likharev, K.~K.} \& \bibinfo{author}{Semenov, V.~K.}
\newblock \bibinfo{title}{{RSFQ logic/memory family: a new Josephson-junction
  technology for sub-terahertz-clock-frequency digital systems}}.
\newblock \emph{\bibinfo{journal}{IEEE Trans. Appl. Supercond.}}
  \textbf{\bibinfo{volume}{1}}, \bibinfo{pages}{3--28} (\bibinfo{year}{1991}).

\bibitem{Yamashita:2012ge}
\bibinfo{author}{Yamashita, T.}, \bibinfo{author}{Miki, S.},
  \bibinfo{author}{Terai, H.}, \bibinfo{author}{Makise, K.} \&
  \bibinfo{author}{Wang, Z.}
\newblock \bibinfo{title}{{Crosstalk-free operation of multielement
  superconducting nanowire single-photon detector array integrated with
  single-flux-quantum circuit in a 0.1~W Gifford{\textendash}McMahon
  cryocooler}}.
\newblock \emph{\bibinfo{journal}{Opt. Lett.}} \textbf{\bibinfo{volume}{37}},
  \bibinfo{pages}{2982--2984} (\bibinfo{year}{2012}).

\bibitem{Migdall:2002}
\bibinfo{author}{Migdall, A.~L.}, \bibinfo{author}{Branning, D.},
  \bibinfo{author}{Castelletto, S.} \& \bibinfo{author}{Ware, M.}
\newblock \bibinfo{title}{{Tailoring single-photon and multiphoton
  probabilities of a single-photon on-demand source}}.
\newblock \emph{\bibinfo{journal}{PRA}} \textbf{\bibinfo{volume}{66}},
  \bibinfo{pages}{053805} (\bibinfo{year}{2002}).

\bibitem{Kuno:2015by}
\bibinfo{author}{Kuno, Y.} \emph{et~al.}
\newblock \bibinfo{title}{{Design of apodized hydrogenated amorphous silicon
  grating couplers with metal mirrors for inter-layer signal coupling: Toward
  three-dimensional optical interconnection}}.
\newblock \emph{\bibinfo{journal}{Japanese Journal of Applied Physics}}
  \textbf{\bibinfo{volume}{54}}, \bibinfo{pages}{04DG04}
  (\bibinfo{year}{2015}).

\bibitem{Zhang:2014hn}
\bibinfo{author}{Zhang, H.} \emph{et~al.}
\newblock \bibinfo{title}{{Efficient silicon nitride grating coupler with
  distributed Bragg reflectors}}.
\newblock \emph{\bibinfo{journal}{Op. Ex.}} \textbf{\bibinfo{volume}{22}},
  \bibinfo{pages}{21800--21805} (\bibinfo{year}{2014}).

\bibitem{Pelc:2014hz}
\bibinfo{author}{Pelc, J.~S.} \emph{et~al.}
\newblock \bibinfo{title}{{Picosecond all-optical switching in hydrogenated
  amorphous silicon microring resonators}}.
\newblock \emph{\bibinfo{journal}{Op. Ex.}} \textbf{\bibinfo{volume}{22}},
  \bibinfo{pages}{3797} (\bibinfo{year}{2014}).

\bibitem{Rambo:ZC8H7tI3}
\bibinfo{author}{Rambo, T.~M.}, \bibinfo{author}{McCusker, K.~T.},
  \bibinfo{author}{Huang, Y.-P.} \& \bibinfo{author}{Kumar, P.}
\newblock \bibinfo{title}{{Low-loss all-optical quantum switching}}.
\newblock In \emph{\bibinfo{booktitle}{2013 IEEE Photonics Society Summer
  Topical Meeting Series}}, \bibinfo{pages}{WD2.3--180}
  (\bibinfo{publisher}{IEEE}, \bibinfo{year}{2013}).

\bibitem{Turner:2008bc}
\bibinfo{author}{Turner, A.~C.}, \bibinfo{author}{Foster, M.~A.},
  \bibinfo{author}{Gaeta, A.~L.} \& \bibinfo{author}{Lipson, M.}
\newblock \bibinfo{title}{{Ultra-low power parametric frequency conversion in a
  silicon microring resonator}}.
\newblock \emph{\bibinfo{journal}{Op. Ex.}} \textbf{\bibinfo{volume}{16}},
  \bibinfo{pages}{4881--4887} (\bibinfo{year}{2008}).

\bibitem{Rong:2005bb}
\bibinfo{author}{Rong, H.} \emph{et~al.}
\newblock \bibinfo{title}{{An all-silicon Raman laser}}.
\newblock \emph{\bibinfo{journal}{Nature}} \textbf{\bibinfo{volume}{433}},
  \bibinfo{pages}{292--294} (\bibinfo{year}{2005}).

\bibitem{Liu:2010by}
\bibinfo{author}{Liu, X.}, \bibinfo{author}{Osgood, R.~M.},
  \bibinfo{author}{Vlasov, Y.~A.} \& \bibinfo{author}{Green, W. M.~J.}
\newblock \bibinfo{title}{{Mid-infrared optical parametric amplifier using
  silicon nanophotonic waveguides}}.
\newblock \emph{\bibinfo{journal}{Nature Photonics}}
  \textbf{\bibinfo{volume}{4}}, \bibinfo{pages}{557--560}
  (\bibinfo{year}{2010}).

\bibitem{Espinola:2004kv}
\bibinfo{author}{Espinola, R.}, \bibinfo{author}{Dadap, J.},
  \bibinfo{author}{Richard~Osgood, J.}, \bibinfo{author}{McNab, S.} \&
  \bibinfo{author}{Vlasov, Y.}
\newblock \bibinfo{title}{{Raman amplification in ultrasmall
  silicon-on-insulator wire waveguides}}.
\newblock \emph{\bibinfo{journal}{Op. Ex.}} \textbf{\bibinfo{volume}{12}},
  \bibinfo{pages}{3713--3718} (\bibinfo{year}{2004}).

\bibitem{Leuthold:2009gq}
\bibinfo{author}{Leuthold, J.} \emph{et~al.}
\newblock \bibinfo{title}{{Silicon Organic Hybrid Technology---A Platform for
  Practical Nonlinear Optics}}.
\newblock \emph{\bibinfo{journal}{Proceedings of the IEEE}}
  \textbf{\bibinfo{volume}{97}}, \bibinfo{pages}{1304--1316}
  (\bibinfo{year}{2009}).

\bibitem{Xiong:11}
\bibinfo{author}{Xiong, C.} \emph{et~al.}
\newblock \bibinfo{title}{Integrated gan photonic circuits on silicon (100) for
  second harmonic generation}.
\newblock \emph{\bibinfo{journal}{Opt. Express}} \textbf{\bibinfo{volume}{19}},
  \bibinfo{pages}{10462--10470} (\bibinfo{year}{2011}).

\bibitem{Wang:2013jg}
\bibinfo{author}{Wang, T.} \emph{et~al.}
\newblock \bibinfo{title}{{Multi-photon absorption and third-order nonlinearity
  in silicon at mid-infrared wavelengths}}.
\newblock \emph{\bibinfo{journal}{Op. Ex.}} \textbf{\bibinfo{volume}{21}},
  \bibinfo{pages}{32192--32198} (\bibinfo{year}{2013}).

\bibitem{Jalali:2010fo}
\bibinfo{author}{Jalali, B.}
\newblock \bibinfo{title}{{Silicon photonics: Nonlinear optics in the
  mid-infrared}}.
\newblock \emph{\bibinfo{journal}{Nature Photonics}}
  \textbf{\bibinfo{volume}{4}}, \bibinfo{pages}{506--508}
  (\bibinfo{year}{2010}).

\bibitem{Marsili:2013hy}
\bibinfo{author}{Marsili, F.} \emph{et~al.}
\newblock \bibinfo{title}{{Mid-Infrared Single-Photon Detection with Tungsten
  Silicide Superconducting Nanowires}}.
\newblock \emph{\bibinfo{journal}{CLEO: 2013 (2013), paper CTu1H.1}}
  \bibinfo{pages}{CTu1H.1} (\bibinfo{year}{2013}).

\bibitem{Baek:2011cs}
\bibinfo{author}{Baek, B.}, \bibinfo{author}{Lita, A.~E.},
  \bibinfo{author}{Verma, V.} \& \bibinfo{author}{Nam, S.~W.}
\newblock \bibinfo{title}{{Superconducting a-W x Si 1 x nanowire single-photon
  detector with saturated internal quantum efficiency from visible to 1850
  nm}}.
\newblock \emph{\bibinfo{journal}{Applied Physics Letters}}
  \textbf{\bibinfo{volume}{98}}, \bibinfo{pages}{251105}
  (\bibinfo{year}{2011}).

\bibitem{Korneeva:2011de}
\bibinfo{author}{Korneeva, Y.} \emph{et~al.}
\newblock \bibinfo{title}{{New Generation of Nanowire NbN Superconducting
  Single-Photon Detector for Mid-Infrared}}.
\newblock \emph{\bibinfo{journal}{IEEE Trans. Appl. Supercond.}}
  \textbf{\bibinfo{volume}{21}}, \bibinfo{pages}{323--326}
  (\bibinfo{year}{2011}).

\bibitem{Gershenzon:1990mm}
\bibinfo{author}{Gershenzon, E.~M.} \emph{et~al.}
\newblock \bibinfo{title}{{Millimeter and submillimeter range mixer based on
  electron heating of superconducting films in the resistive state}}.
\newblock \emph{\bibinfo{journal}{Sov. Phys. Superconductivity}}
  \textbf{\bibinfo{volume}{3}}, \bibinfo{pages}{1583--1597}
  (\bibinfo{year}{1990}).

\bibitem{Chrostowski:2015bk}
\bibinfo{author}{Chrostowski, L.} \& \bibinfo{author}{Hochberg, M.}
\newblock \emph{\bibinfo{title}{{Silicon Photonics Design}}}.
\newblock From Devices to Systems (\bibinfo{publisher}{Cambridge University
  Press}, \bibinfo{address}{Cambridge}, \bibinfo{year}{2015}).

\bibitem{Barrett:2010goa}
\bibinfo{author}{Barrett, S.~D.} \& \bibinfo{author}{Stace, T.~M.}
\newblock \bibinfo{title}{{Fault Tolerant Quantum Computation with Very High
  Threshold for Loss Errors}}.
\newblock \emph{\bibinfo{journal}{Phys. Rev. Lett.}}
  \textbf{\bibinfo{volume}{105}}, \bibinfo{pages}{200502}
  (\bibinfo{year}{2010}).

\end{thebibliography}
\bibliographystyle{naturemag}




%
\begin{IEEEbiography}[{\includegraphics[width=1in,height=1.25in,clip,keepaspectratio]{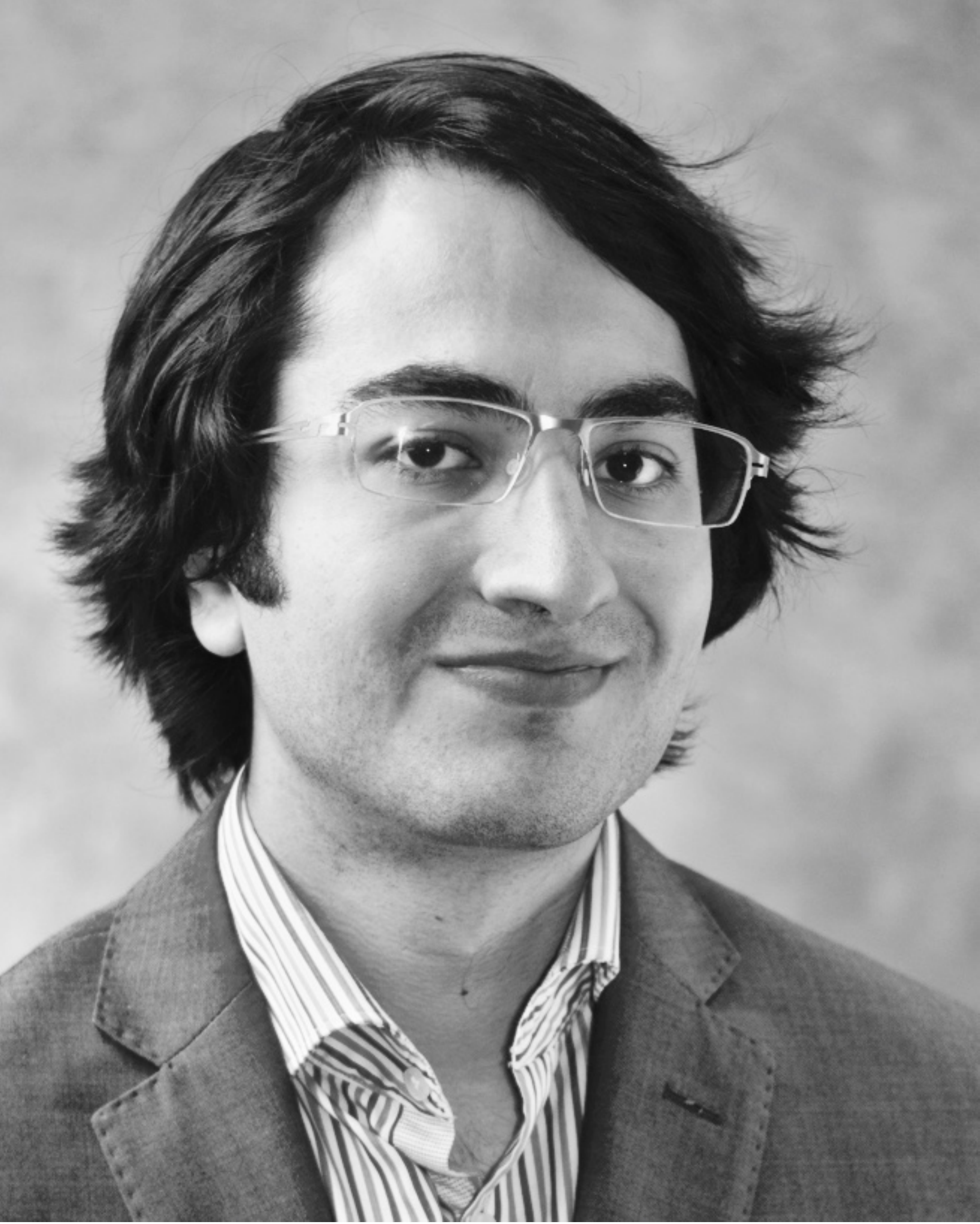}}]{Joshua W. Silverstone}
received his BSc in engineering physics from the University of Alberta, Canada. There, he developed biosensors based on silicon quantum dots in whispering gallery mode cavities, as well as novel silicon-based integrated optics. He completed his PhD at the University of Bristol, UK, in 2015, pioneering the nascent field of quantum optics in silicon waveguides. He is now Postdoctoral Research Associate at the University of Bristol, developing quantum applications using large-scale integrated optics technology.
\end{IEEEbiography}

\begin{IEEEbiography}[{\includegraphics[width=1in,height=1.25in,clip,keepaspectratio]{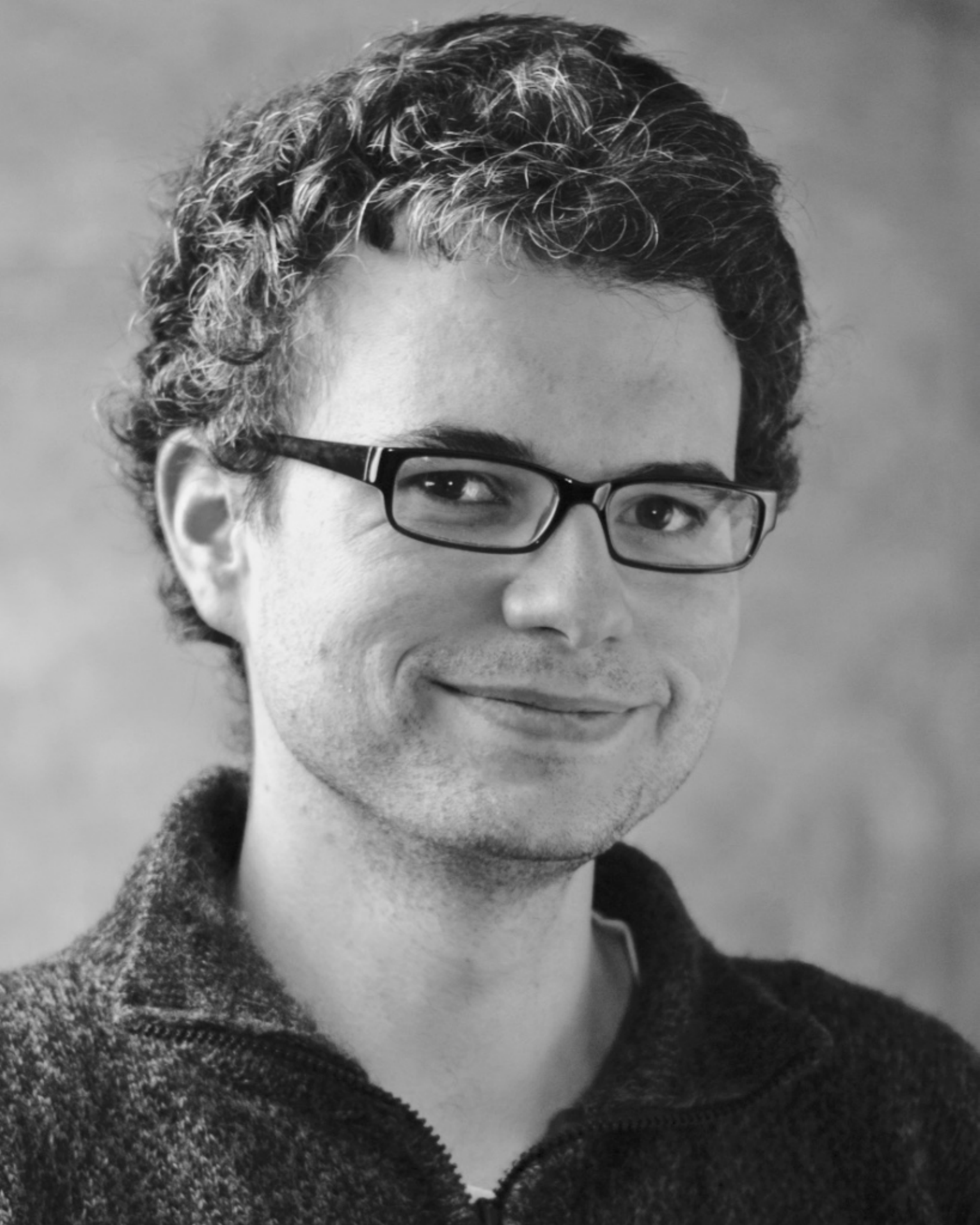}}]{Damien Bonneau}
received both his dipl\^{o}me d'ing\'enieur from T\'el\'ecom Physique Strasbourg and MSc in particle physics from Universit\'e Louis Pasteur, Strasbourg, France in 2007. He received the Ph. D. degree in physics from the University of Bristol, UK in 2014 for his work on silicon photonic quantum circuits at the Centre for Quantum Photonics. He is now a Postdoctoral Research Associate in CQP. His research focuses mainly on silicon quantum photonics, including both component- and system-level design and characterization.
\end{IEEEbiography}


\begin{IEEEbiography}[{\includegraphics[width=1in,height=1.25in,clip,keepaspectratio]{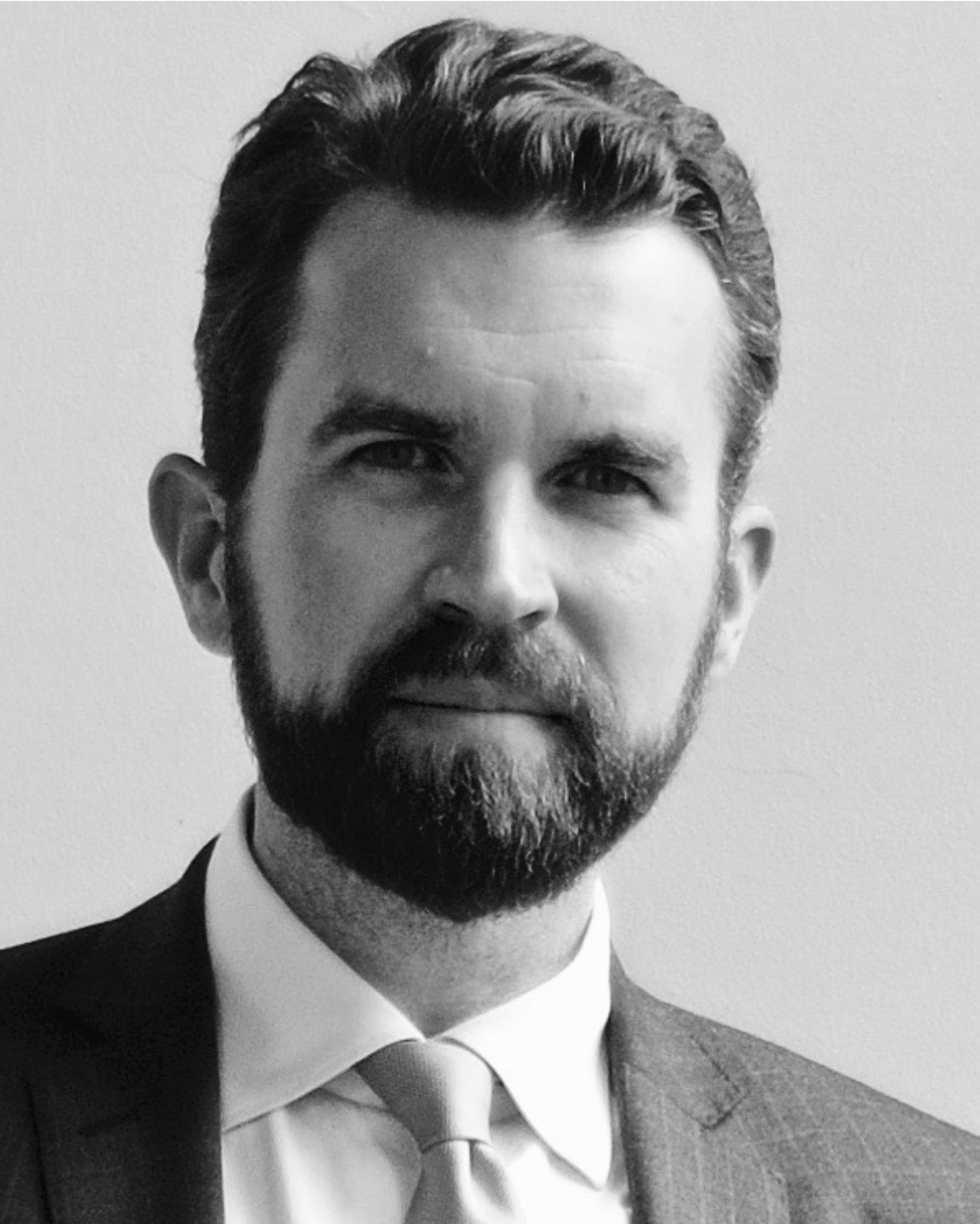}}]{Jeremy L. O'Brien}
 is the director of the Centre for Quantum Photonics at the University of Bristol, where he is Professor in Physics and Electrical Engineering. He received his PhD from the University of New South Wales in 2002 for experimental work on correlated and confined electrons in super- and semi-conductor nanostructures, and on quantum computation with phosphorus-in-silicon qubits. As a research fellow at the University of Queensland (2001-2006) he developed quantum optics and quantum information science with single photons.
\end{IEEEbiography}

\begin{IEEEbiography}[{\includegraphics[width=1in,height=1.25in,clip,keepaspectratio]{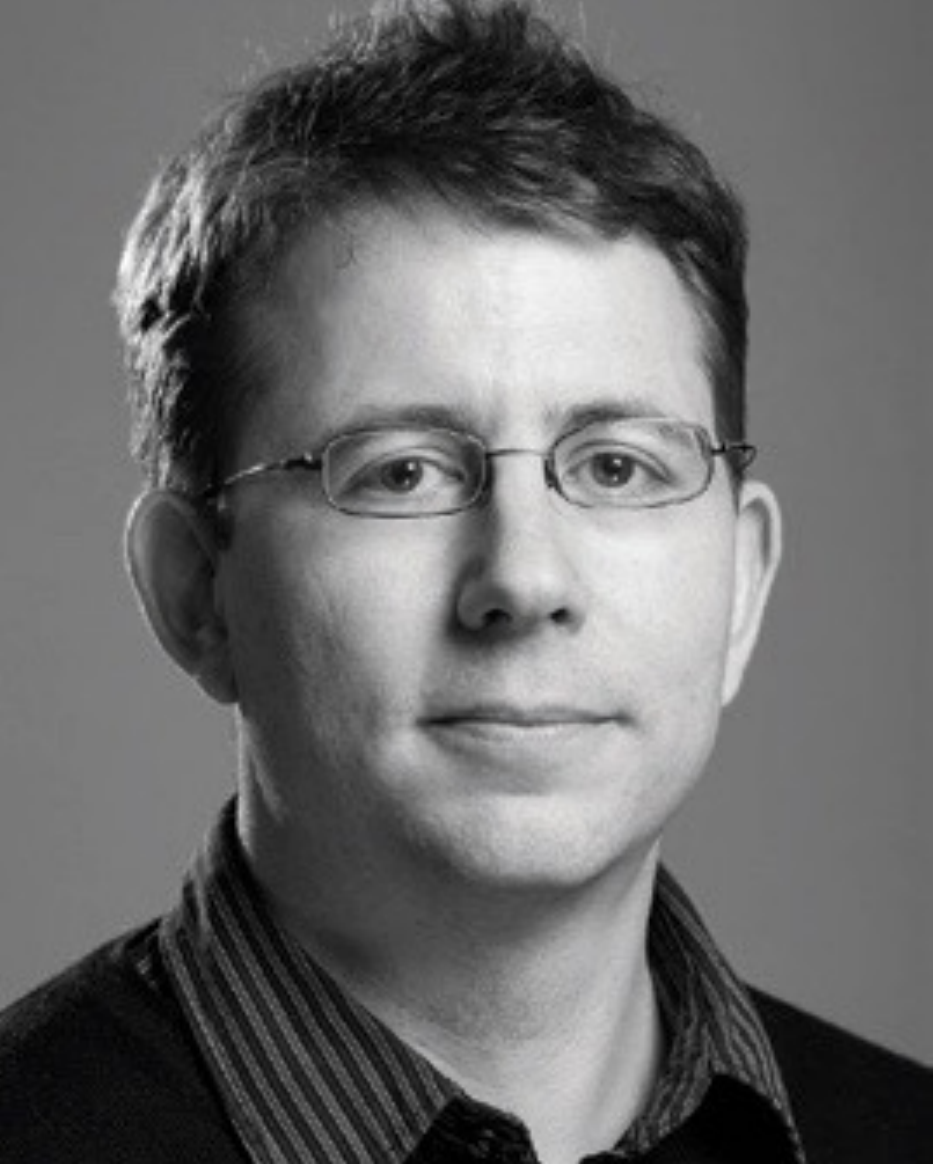}}]{Mark G. Thompson}
is Professor of Quantum Photonics at the University of Bristol UK, and Director of the Centre for Doctoral Training in Quantum Engineering. He received his M.Phys. degree in Physics from the University of Sheffield UK, and  Ph.D. degree in Electrical Engineering from the University of Cambridge UK, in 2000 and 2007, respectively. He was a Research Scientist with Bookham Technology Ltd. UK, a Research Fellow at both the University of Cambridge and the Toshiba Corporate Research \& Development Centre in Kawasaki, Japan. His research interests are in quanutm photonic devices and applications.
\end{IEEEbiography}




\end{document}